\documentclass[preprint,3p,12pt]{elsarticle}
\pdfoutput=1
\usepackage{amsmath}
\usepackage{amssymb}
\usepackage{booktabs}
\usepackage{multirow}
\usepackage{float}
\usepackage{subfigure}
\usepackage{lineno,hyperref}
\usepackage{epstopdf}
\usepackage[noend]{algpseudocode}
\usepackage{algorithmicx,algorithm}

\modulolinenumbers[5]
\begin{document}
\begin{frontmatter}
\title{An efficient targeted ENO scheme with local adaptive dissipation for compressible flow simulation}

\author[IMECH,UCAS]{Jun Peng\corref{cor}}
\ead{pengjun@imech.ac.cn}
\author[IMECH,UCAS,IAPCM]{Shiyao Li}
\author[IMECH,UCAS]{Yiqing Shen\corref{cor}}
\ead{yqshen@imech.ac.cn}
\author[IAPCM]{Shengping Liu}
\author[IMECH,UCAS]{Ke Zhang}

\cortext[cor]{Corresponding author}
\address[IMECH]{State Key Laboratory of High Temperature Gas Dynamics, Institute of Mechanics, Chinese Academy of Sciences, Beijing, China}
\address[UCAS]{School of Engineering Science, University of Chinese Academy of Sciences, Beijing, China}
\address[IAPCM]{Institute of Applied Physics and Computational Mathematics, Beijing 100094, China}

\begin{abstract}
High fidelity numerical simulation of compressible flow requires the numerical method being used to have both stable shock-capturing capability and high spectral resolution. Recently, a family of Targeted Essentially Non-Oscillatory (TENO) schemes are developed to fulfill such requirements. Although TENO has very low dissipation for smooth flow, it introduces a cutoff value $C_T$ to maintain the non-oscillatory shock-capturing property. $C_T$ is problem depended and therefore needs adjustments by trial and error for different problems. As $C_T$ actually controls the dissipation property of TENO, the choice of $C_T$ for better shock-capturing capability always means higher dissipation. To overcome this, in this paper, a local adaptive method is proposed for the choice of $C_T$. By introducing a novel adaptive function based on the WENO smoothness indicators, $C_T$ is dynamically adjusted form $1.0 \times 10.0^{-10}$ for lower dissipation to $1.0 \times 10.0^{-4}$ for better capturing of shock according to the smoothness of the reconstruction stencil.  Numerical results of the new method are compared with those of the original TENO method and the TENO-A method (Fu et al., JCP, 2017). It reveals that the new method is capable of better suppressing numerical oscillations near discontinuities while maintaining the low dissipation property of TENO at lower extra computational cost.
\end{abstract}

\end{frontmatter}

\section{Introduction}\label{sec1}
Compressible flow is ubiquitous in engineering and scientific researches. It is featured by multi-scale spacial/temporal structures like turbulence, discontinuities like shockwaves, and the interactions of such structures. Accurate and high fidelity numerical simulation of compressible flow requires numerical methods that are capable of simultaneously resolving all of these flow phenomena. However, different structures have different demands: low-dissipation for multi-scale structures and high-dissipation for discontinuities. Such contradictory requirements imply that a numerical scheme should be able to adjust its dissipation property in accord to the flow structure being locally resolved.

The state-of-the-art study of numerical methods for compressible flow concerns the developing of high-order shock-capturing schemes. Such methods suppress numerical oscillations in vicinity of discontinuities by increasing dissipation when local gradient is large enough while maintaining low dissipation in smooth regions. 

By using intrinsic limiting procedures, classical high-order shock-capturing schemes, as exemplified by total variation diminishing (TVD) schemes \cite{harten1983high}, essentially non-oscillatory (ENO) schemes \cite{harten1987uniformly}, and so on \cite{Pirozzoli2011}, can resolve discontinuities without numerical oscillations, nevertheless, they suffer from excessive numerical dissipation in smooth regions. The family of weighted essentially non-oscillatory (WENO) finite difference schemes \cite{liu1994weighted,jiang1996efficient} achieves lower dissipation for smooth fields as well as non-oscillatory shock-capturing capability. Based on the smoothness of each sub-stencil, WENO schemes dynamically adjust local numerical dissipation in an elaborate weighting approach. Within the general framework of the WENO-JS scheme by Jiang \& Shu \cite{jiang1996efficient}, the WENO family is further extended and developed. Henrick et al. \cite{henrick2005mapped} find that the WENO-JS scheme, at critical points, does not satisfy the fifth order convergence conditions and the WENO-M scheme is proposed to circumvent such problem by mapping of the WENO-JS weights. Borges et al. \cite{borges2008improved} introduce a higher order smoothness indicator for the non-linear weights and propose the WENO-Z scheme which satisfies the sufficient criteria for fifth-order convergence at much lower computational cost than the WENO-M method. The accuracy of the WENO-Z scheme is further improved in \cite{Castro:2011cz,DON2013347,Yamaleev:2009hs,Yamaleev:2009gp,Fan:2014kj,Liu2018}. Shen and Zha \cite{Shen:2014gf} show that at transitional points, which connect smooth region and discontinuity, the accuracy of fifth order WENO schemes is second order and a series of multi-step weighting methods \cite{shen:2014gfa,Peng:2015hn,ma2016,Zeng2019} are developed to improve the accuracy. Other high-order methods are also developed based on the ideal of WENO in \cite{Zhu2016,Qiu2004,Qiu2005,he2015preventing,levy_puppo_russo_1999,cravero2018cweno,CRAVERO2017}. Even though the order of accuracy for WENO schemes can be designed to be arbitrarily high \cite{martin2006bandwidth,Hu2010,balsara2000monotonicity,Gerolymos:2009dn}, the spectral resolution of WENO schemes is still not satisfactory \cite{pirozzoli2006spectral}. Specifically, their excessive numerical dissipation for small scale structures in compressible turbulent flows may overwhelm physical dissipation \cite{johnsen2010assessment}. Such high dissipation, compared to linear schemes of the same order, is related to the weighting mechanism and the smoothness indicators of the WENO scheme.

By using a switcher/shock-sensor to classify the flow field as smooth or discontinuous, hybrid schemes \cite{adams1996high,pirozzoli2002conservative,hill2004hybrid,Costa:2007dl} apply low dissipation schemes such as compact schemes \cite{lele1992compact} in smooth regions and shock capturing schemes such as the WENO scheme when discontinuities emerge. The high dissipation caused by the weighting mechanism of WENO for smooth region is thus avoided. Varieties of methods have been proposed to develop switcher/shock-sensor method \cite{ZHAO2020104439,Li2010}, such as the flow variable difference between neighboring points \cite{adams1996high,pirozzoli2002conservative,hill2004hybrid}, the flow gradients \cite{johnsen2010assessment,Ziegler:2011wz}, the multi-resolution coefficients \cite{Costa:2007dl,Gao:2012jg}, and the Boundary Value Diminishing (BVD) criterion \cite{Deng2019,Sun2016}. Obviously, the performance of a hybrid scheme relies on the accuracy of its switcher/shock-sensor. 

Beyond typical WENO and hybrid approaches, an novel way to control numerical dissipation is introduced in the family of targeted ENO (TENO) schemes \cite{fu2016family,fu2017,fu2018}. By using a scale separation technique \cite{Hu2011}, TENO cuts off the least smooth sub-stencil and therefore avoids the high dissipation issue caused by the WENO weights. The idea of TENO can be extended to arbitrary high order \cite{Fu2019,zhang2018extending}. TENO achieves significant low dissipation for both smooth, however, it still produces numerical oscillations for some strong discontinuities. This indicates that the numerical dissipation for discontinuities of TENO is not enough. The dissipation property of TENO is strongly related to the cut of value $C_T$ which is problem dependent and requires tuning. Based on a non-linear shock detector, an adaptive version of $C_T$ is developed in \cite{fu2018}. This adaptive TENO scheme increases $C_T$ near discontinuities for non-oscillatory shock-capturing and decreases it in smooth regions for high resolution. However, as a set of extra free parameters are introduced, deeper tuning process is required to obtain balanced robustness and low dissipation. 

In practice, parameter tuning is generally time consuming especially for large scale numerical simulations like the DNS (Direct Numerical Simulation) of supersonic turbulent flow. In this paper, we propose a new adaptive method to improve the performance of TENO. Based on the shock-sensor developed in \cite{peng2017novel}, an novel adaptive function for $C_T$ is introduced. The new method maintains the low dissipation property of TENO for smooth fields and significantly improves its shock-capturing capability at a low price. This paper is organized as following. In \ref{sec2}, the TENO scheme is briefly reviewed. The new method is proposed and analyzed in \ref{sec3}. Numerical validations are presented in \ref{sec4}. Concluding remarks are given in \ref{sec5}.

\section{The fifth order targeted ENO scheme of Fu et al. \cite{fu2016family,fu2018}}\label{sec2}
To describe the TENO scheme, we consider the one-dimensional hyperbolic conservation law expressed as:
\begin{equation}\label{eq:2.1}
\frac{\partial u}{\partial t}+\frac{\partial f}{\partial x}=0
\end{equation}
where $u(x,t)$ is the conserved variable and $f(u)$ is the flux function. To solve \eqref{eq:2.1} numerically, we transform it into semi-discretized form on uniformly discretized space:
\begin{equation}\label{eq:2.2}
\frac{du_i}{dt}=-\frac{\hat{f}_{i+\frac{1}{2}}-\hat{f}_{i-\frac{1}{2}}}{\Delta x}
\end{equation}
in which $u_i=u(x_i)$, $\hat{f}_{i+\frac{1}{2}}=\hat{f}_{i+\frac{1}{2}}^{+}+\hat{f}_{i+\frac{1}{2}}^{-}$ is the numerical flux at cell interface $x_{i+\frac{1}{2}}=x_{i}+\Delta x/2$ and $\Delta x =x_{i+1}-x_{i}$. The splitted numerical fluxes $\hat{f}^\pm_{i+\frac{1}{2}}$ at cell interface are to be reconstructed. For simplicity, $\pm$ in the superscript are dropped in the following parts of this paper.

The numerical flux $\hat{f}_{i+\frac{1}{2}}$ can be obtained by high order schemes. The fifth order upstream-biased scheme is written as:
\begin{equation}\label{eq:2.3}
	\hat{f}_{i+\frac{1}{2}}=\frac{2}{60}{f}_{i-2}-\frac{13}{60}{f}_{i-1}+\frac{47}{60}{f}_{i}+\frac{27}{60}{f}_{i+1}-\frac{3}{60}{f}_{i+2},
\end{equation}
where $f_i=f(u_i)$ is the point value of the flux. Eq.\eqref{eq:2.3} is a convex combination of three third order schemes on three sub-stencils $S_0=(x_{i-2},x_{i-1},x_{i})$, $S_1=(x_{i-1},x_{i},x_{i+1})$, and $S_2=(x_i,x_{i+1},x_{i+2})$:
\begin{align}
\hat{f}_{0,i+1/2} &=\frac{1}{3}{{f}_{i-2}}-\frac{7}{6}{{f}_{i-1}}+\frac{11}{6}{{f}_{i}},\label{eq:2.4}\\
\hat{f}_{1,i+1/2} &=-\frac{1}{6}{{f}_{i-1}}+\frac{5}{6}{{f}_{i}}+\frac{1}{3}{{f}_{i+1}},\label{eq:2.5}\\
\hat{f}_{2,i+1/2} &=\frac{1}{3}{{f}_{i}}+\frac{5}{6}{{f}_{i+1}}-\frac{1}{6}{{f}_{i+2}}.\label{eq:2.6}
\end{align}
with linear weights 
$$
c_0=0.1, \quad c_1=0.6, \quad c_2=0.3
$$ 
respectively. By substituting the linear weights with the non-linear TENO weights, we have the fifth order TENO scheme:
\begin{equation}\label{eq:2.7}
\text{TENO5}: \quad \hat{f}_{i+1/2}={\omega}_{0}\hat{f}_{0,i+1/2}+\omega_{1}\hat{f}_{1,i+1/2}+\omega_{2}\hat{f}_{2,i+1/2}.
\end{equation}

To compute the non-linear weights $\omega_{k}$, the smoothness of each sub-stencil is firstly measured by:
\begin{equation}\label{eq:2.8}
\gamma_{k} =\left(C+\frac{\tau_K}{\beta_k + \epsilon}\right)^q,k=0,1,2
\end{equation}
in which $\tau_K$ is the global smoothness indicator, $\beta_k$ is the WENO smoothness indicator \cite{jiang1996efficient} of each sub-stencil given by:
\begin{align}
{{\beta }_{0}} &=\frac{13}{12}{{({{f}_{i-2}}-2{{f}_{i-1}}+{{f}_{i}})}^{2}}+\frac{1}{4}{{({{f}_{i-2}}-4{{f}_{i-1}}+3{{f}_{i}})}^{2}},\label{eq:2.9}\\
{{\beta }_{1}} &=\frac{13}{12}{{({{f}_{i-1}}-2{{f}_{i}}+{{f}_{i+1}})}^{2}}+\frac{1}{4}{{({{f}_{i-1}}-{{f}_{i+1}})}^{2}},\label{eq:2.10}\\
{{\beta }_{2}} &=\frac{13}{12}{{({{f}_{i}}-2{{f}_{i+1}}+{{f}_{i+2}})}^{2}}+\frac{1}{4}{{(3{{f}_{i}}-4{{f}_{i+1}}+{{f}_{i+2}})}^{2}}.\label{eq:2.11}
\end{align}
Noted that the choice for $\tau_K$ is flexible for TENO \cite{fu2016family}.  The global smoothness indicator of WENO-Z, i.e.: 
\begin{equation}\label{eq:2.12}
\tau_K = \tau_5 = |\beta_2 - \beta_0|
\end{equation}
is used in this paper. 
$C$ and $q$ are parameters to incorporate a scale-separation mechanism \cite{Hu2011} with typical values $C=1$ and $q=6$. $\epsilon$ is a small value to avoid division by zero as in the WENO scheme. $\epsilon$ is set to $1 \times 10^{-6}$ in this paper.

Then, a ENO-like stencil selection method is applied. The smoothness measure Eq.\eqref{eq:2.8} is normalized:
\begin{equation}\label{eq:2.13}
\chi_k = \frac{\gamma_k}{\sum_{k=0}^{2} \gamma_k}, \quad k=0,1,2. 
\end{equation}
and passed to a cutoff function:
\begin{equation}\label{eq:2.14}
\delta_k = \begin{cases}
0 \quad & \chi_k < C_T,\\
1 \quad & \text{otherwise}.
\end{cases}
\end{equation}
The cutoff value $C_T$ requires tuning for different cases, typically $C_T=1 \times 10^{-6}$.

Finally, the non-linear TENO weights is computed as:
\begin{equation}\label{eq:2.15}
\omega_k = \frac{\delta_k c_k}{\sum_{k=0}^{2} \delta_k c_k}, \quad k=0,1,2. 
\end{equation}

The dissipation property of TENO is related to the choice of the cutoff value $C_T$ \cite{fu2016family,fu2017}. Smaller $C_T$ brings lower dissipation but may also lead to numerical oscillations. An adaptive $C_T$ is proposed in \cite{fu2018}:
\begin{equation}\label{eq:2.16}
\begin{cases}
C_T & = 10^{-\lfloor \beta \rfloor},\\
\beta & = \alpha_1 - \alpha_2(1-g(m)),\\
g(m) &= (1-m)^4(1+4m),
\end{cases}
\end{equation}
where
\begin{equation}\label{eq:2.17}
\begin{cases}
m & = 1-min(1,\frac{\eta_{i+1/2}}{C_r})\\
\eta_{i+1/2} & = min(\eta_{i-1},\eta_{i},\eta_{i+1})
\end{cases}
\end{equation}
in which 
\begin{equation}\label{eq:2.18}
\eta_i=\frac{2|\Delta f_{i+1/2}\Delta f_{i-1/2}|+\epsilon}{(\Delta f_{i+1/2})^2+(\Delta f_{i+1/2})^2 + \epsilon}, \quad \Delta f_{i+1/2}=f_{i+1}-f_i, \quad \epsilon = \frac{0.9C_r}{1-0.9C_r}\xi^2
\end{equation}
The values of the parameters are $\alpha_1=10.5$, $\alpha_2=3.5$, $C_r=0.25$, and $\xi=10^{-3}$. $\lfloor \cdot \rfloor$ denotes the floor function. The setup \eqref{eq:2.16} gives $C_T$ a dynamically adjustable interval according to the smoothness of the flow field ranging from $10^{-7}$ for shock-capturing to $10^{-10}$ for low dissipation. The TENO scheme with adaptive $C_T$ Eq.\eqref{eq:2.16} is referred to as TENO5-A.

\section{The new method}\label{sec3}
The adaptive method \eqref{eq:2.16} for $C_T$ brings extra computational cost. Besides, more parameters requiring further tuning ($C_r$, $\alpha_1$, $\alpha_2$, and $\xi$) are introduced. Also, it has been shown that the upper bound $10^{-7}$ for $C_T$ is still insufficient for robust shock-capturing, larger value of $C_T$ is required e.g. $10^{-4}$ \cite{fleischmann2019}.

To design a more efficient adaptive $C_T$, we propose following principles:
\begin{itemize}
\item\label{rule1} The adaptive method only uses informations that have already been provided by TENO, e.g. $\beta_k$, $\tau_K$. 
\item\label{rule2} The upper and lower bounds of $C_T$ are given by magnitudes instead of values.
\item\label{rule3} The fewer free parameters the better.
\end{itemize}

Following these principles, we propose a new adaptive $C_T$:
\begin{equation}\label{eq:3.1}
\begin{cases}
C_T & = 10^{-m},\\
m &= B_{l} + \lfloor \theta (B_u - B_l)\rfloor,
\end{cases}
\end{equation}
where
\begin{equation}\label{eq:3.2}
\theta = \frac{1}{1+\left(\max_{k}{\tilde{\chi}_k}/H\right)}, \quad \tilde{\chi}_k =\frac{\tau_K}{\beta_k + \epsilon}, \quad k=0,1,2.
\end{equation}
$\theta$ is the Runge function that it has been used as shock-sensor \cite{Peng:2015hn} and switcher \cite{Peng2019} to design adaptive methods. H is a intensity threshold for discontinuities. For example, by taking $H=10$, it means that if the maximum of $\tilde{\chi}_k$ is over 10 then the whole stencil is considered to be discontinuous. Parameters $B_l$ and $B_u$ are the lower magnitude and upper magnitudes of $C_T$. Typical values of $B_l$ and $B_u$ are:
$$
B_l = 4, \quad B_u = 10.
$$
According the smoothness of the stencil, $\theta$ varies from 0 for discontinuity to 1 for smooth field. The value of $C_T$ therefore varies from $10^{-B_l}$ to $10^{-B_u}$. Noted that $C_T$ equals $10^{-B_u}$ only when $\theta$ is exactly $1.0$, i.e. $\max_{k}{\tilde{\chi}_k}/H = 0.0$.

Considering that the computational cost for computing $10^{-m}$ on computer is very high. We introduce a constant array, the Ladder array, to store pre-calculated $C_T$s of different magnitudes:
\begin{equation}\label{eq:3.3}
\text{Ladder}: \quad Lad(B_l:B_u) = (10^{-B_l},...,10^{-B_u}).
\end{equation}
$C_T$ is then determined by:
\begin{equation}\label{eq:3.4}
C_T = Lad(m).
\end{equation}
In the following part of the paper, we refer to the new method as TENO5-LAD (Local Adaptive Dissipation)

\begin{figure}
\begin{center}
\subfigure[Dissipation]{
\includegraphics[width=0.9\textwidth]{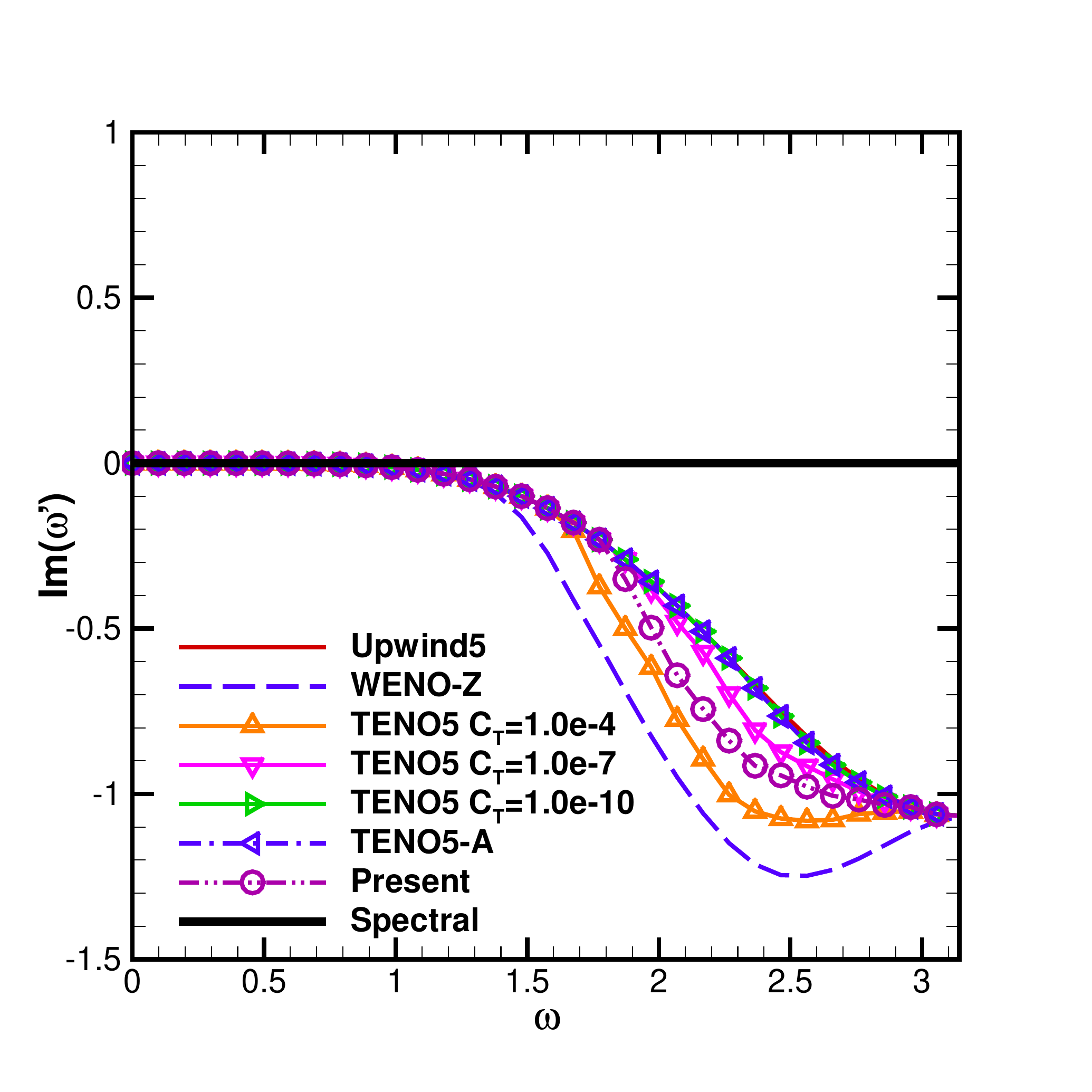}}
\caption{Dispersion and dissipation properties of different schemes.}
\label{fig:1}
\end{center}
\end{figure}

Spectral properties of TENO5-LAD as well as the linear fifth order upwind scheme, the WENO-Z scheme, and TENO schemes with different $C_T$ are illustrated in Fig.\ref{fig:1}. The new scheme maintains good spectral property of the TENO scheme. It is worth noting that although TENO5 and TENO5-A show lower dissipation and dispersion error at high wave number, they are unable to suppress numerical oscillations as will be shown in \ref{sec4}.

\begin{figure}
\begin{center}
\subfigure[Dispersion]{
\includegraphics[width=0.9\textwidth]{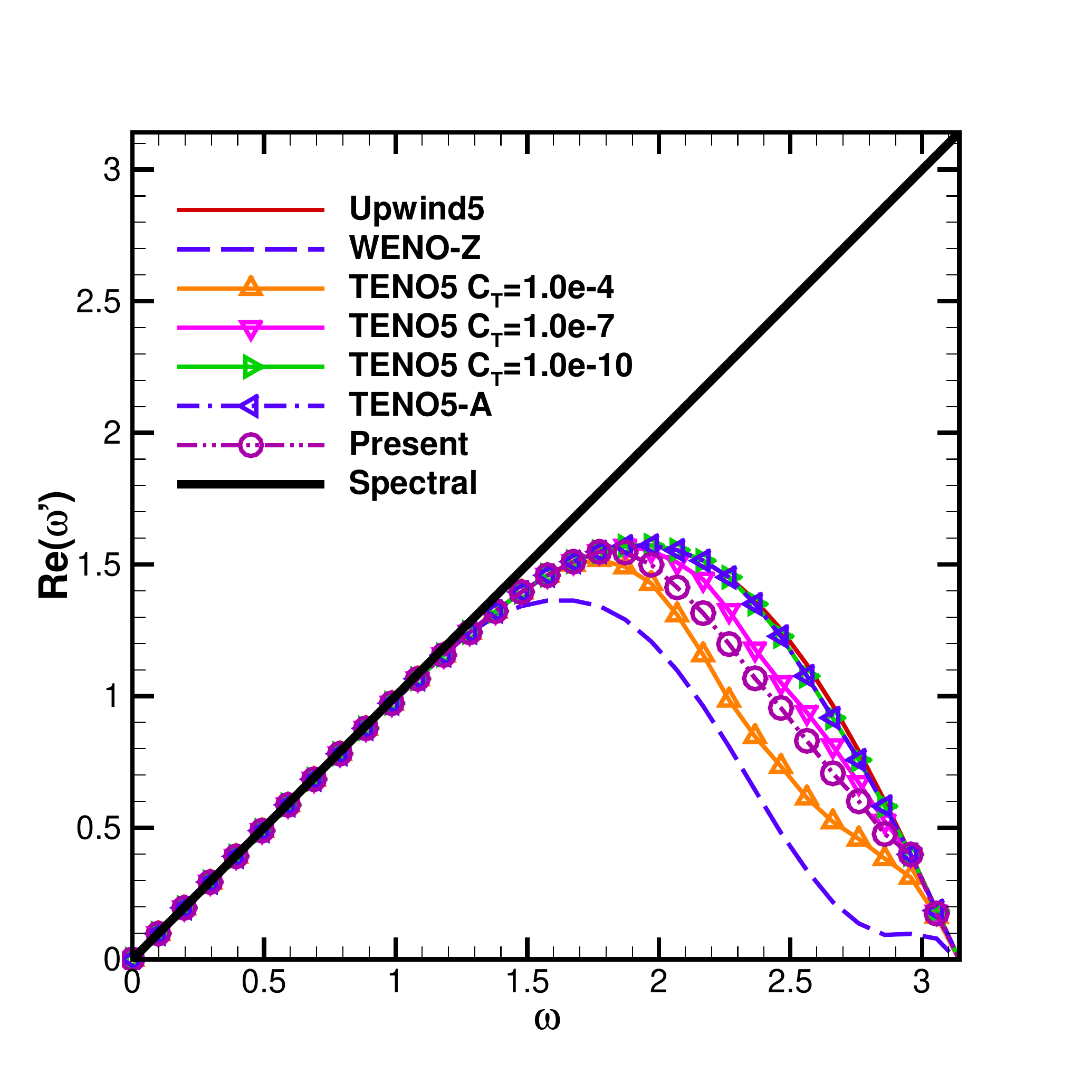}}
\end{center}
\end{figure}

\section{Numerical Validation}\label{sec4}
To assess the new method, we perform several numerical tests including 1D scalar, 1D Euler, and 2D Euler problems. Numerical results are compared with TENO schemes with different $C_t$ setups as well as WENO-Z. Unless specified, $C_T$ for TENO5 is set to be $10^{-7}$, the parameters of the TANO5-A scheme is taken as: 
$$
a_1 = 10.5, \quad a_2 = 3.5, \quad C_r=0.25, \quad \xi = 10^{-3},
$$
and the parameters of the presented method is set to be:
$$
H = 10, \quad B_l = 4, \quad B_u=10.$$

For all of the numerical tests in this section, the third order TVD Runge-Kutta method \cite{shu1988total} is used for time advancing:
\begin{align}
&u^{(1)}=u^n+\Delta tL(u^n)\label{eq:4.1}\\
&u^{(2)}=\frac{3}{4}u^n++\frac{1}{4}u^{(1)}+\frac{1}{4}\Delta tL(u^{(1)})\label{eq:4.2}\\
&u^{n+1}=\frac{1}{3}u^n+\frac{2}{3}u^{(2)}+\frac{2}{3}\Delta tL(u^{(2)})\label{eq:4.3}.
\end{align}
Unless specified, the time step $\Delta t $ is given by:
\begin{equation}\label{eq.dt1}
\Delta t = \sigma \frac{\Delta x}{\max\limits_{i}(|u_i|+\alpha_{i})}
\end{equation}
for one dimensional cases and
\begin{equation}\label{eq.dt2}
\Delta t = \sigma\frac{\Delta t_x \Delta t_y}{\Delta t_x+ \Delta t_y},\ \Delta t_x=\frac{\Delta x}{\max\limits_{i,j}(|u_{i,j}|+\alpha_{i,j})},\ \Delta t_y=\frac{\Delta y}{\max\limits_{i,j}(|v_{i,j}|+\alpha_{i,j})}
\end{equation}
for two dimensional cases, where $\sigma$ is the Courant-Friedrichs-Lewy number.

For the convective terms, the global Lax-Friedrichs splitting method \cite{lax1954weak} is used and the reconstruction of the numerical flux is performed in the characteristic space \cite{ren2003characteristic}.

\subsection{Linear advection equation}\label{sec4.1}
Let us consider the linear wave advection problem. The linear advection equation is given by:
\begin{align}
\begin{cases}
u_t+u_x=0 \quad & -1\leqslant x\leqslant1,\\
u(x,0)=u_0(x) \quad & \text{periodic boundary}.
\end{cases}\label{eq:4.4}
\end{align}
The exact solution of Eq.\eqref{eq:4.4} at time $t$ with the initial condition $u_0(x)$ is given by
\begin{equation}
u(x,t)=u_0(x-t).
\end{equation}
Two cases are studied in this section. 

The first case is to evaluate the convergence order of the present scheme for a smooth solution. The initial condition is given by:
\begin{align}
u_0(x)=sin(\pi x-\frac{sin \pi x}{\pi}).\label{eq:4.5}
\end{align}
This initial condition has two critical points where $f'=0$ and $f'''\neq 0$ \cite{henrick2005mapped}. The time step $\Delta t$ is set to $\Delta x^{5/3}$.

\begin{table}[H]
  \caption{$L_2$ errors and convergence orders for different schemes for the linear advection equation with initial condition \eqref{eq:4.5} at t=2.}\label{tab:1}
  \begin{center}\footnotesize
  \begin{tabular}{ccccccccc}
  \toprule
  \multirow{2}{*}{N} &
  \multicolumn{2}{c}{Upwind5} &
  \multicolumn{2}{c}{TENO5} &
  \multicolumn{2}{c}{TENO5-A} &
  \multicolumn{2}{c}{Present} \\
  & $L_2$ & order& $L_2$ & order & $L_2$ & order & $L_2$ & order \\
  \midrule
  20  & 2.7611E-003 & -    & 2.7611E-003 & -    & 2.7611E-003 & -    & 2.7611E-006 & -    \\
  40  & 9.5732E-005 & 4.85 & 9.5732E-004 & 4.85 & 9.5732E-004 & 4.85 & 9.5732E-006 & 4.85 \\
  80  & 3.0514E-006 & 4.97 & 3.0514E-006 & 4.97 & 3.0514E-006 & 4.97 & 3.0514E-006 & 4.97 \\
  160 & 9.6010E-008 & 4.99 & 9.6010E-007 & 4.99 & 9.6010E-007 & 4.99 & 9.6010E-006 & 4.99 \\
  320 & 3.0061E-009 & 5.00 & 3.0061E-009 & 5.00 & 3.0061E-009 & 5.00 & 3.0061E-006 & 5.00 \\
  \bottomrule
  \end{tabular}
  \end{center}
\end{table}
The $L_2$ norm of the error is obtained by comparison with the exact solution at $t=2$ according to:
$$
L_2=\sqrt{\frac{1}{N}\sum_{i=1}^N{\left(u_i-u_{exact,i}\right)^2}}
$$
Tab.\ref{tab:1} shows the $L_2$ norms as well as convergence orders for different schemes. The differences of $L_2$ errors between different schemes are trivial. 

\begin{figure}[H]
  \begin{center}
  \subfigure[The Gaussian wave]{
  \includegraphics[width=0.45\textwidth]{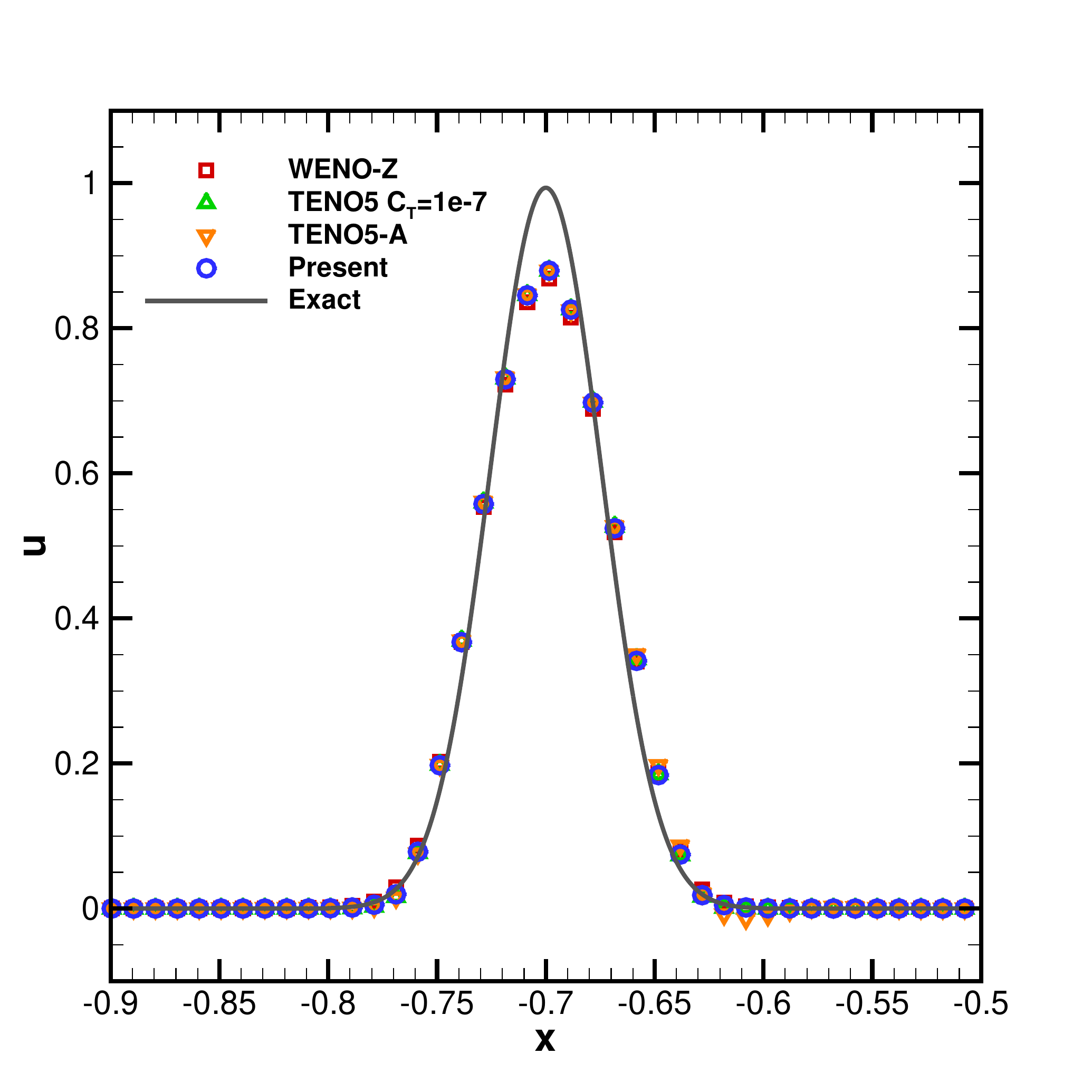}}\label{fig2:a}
  \subfigure[The square wave]{
  \includegraphics[width=0.45\textwidth]{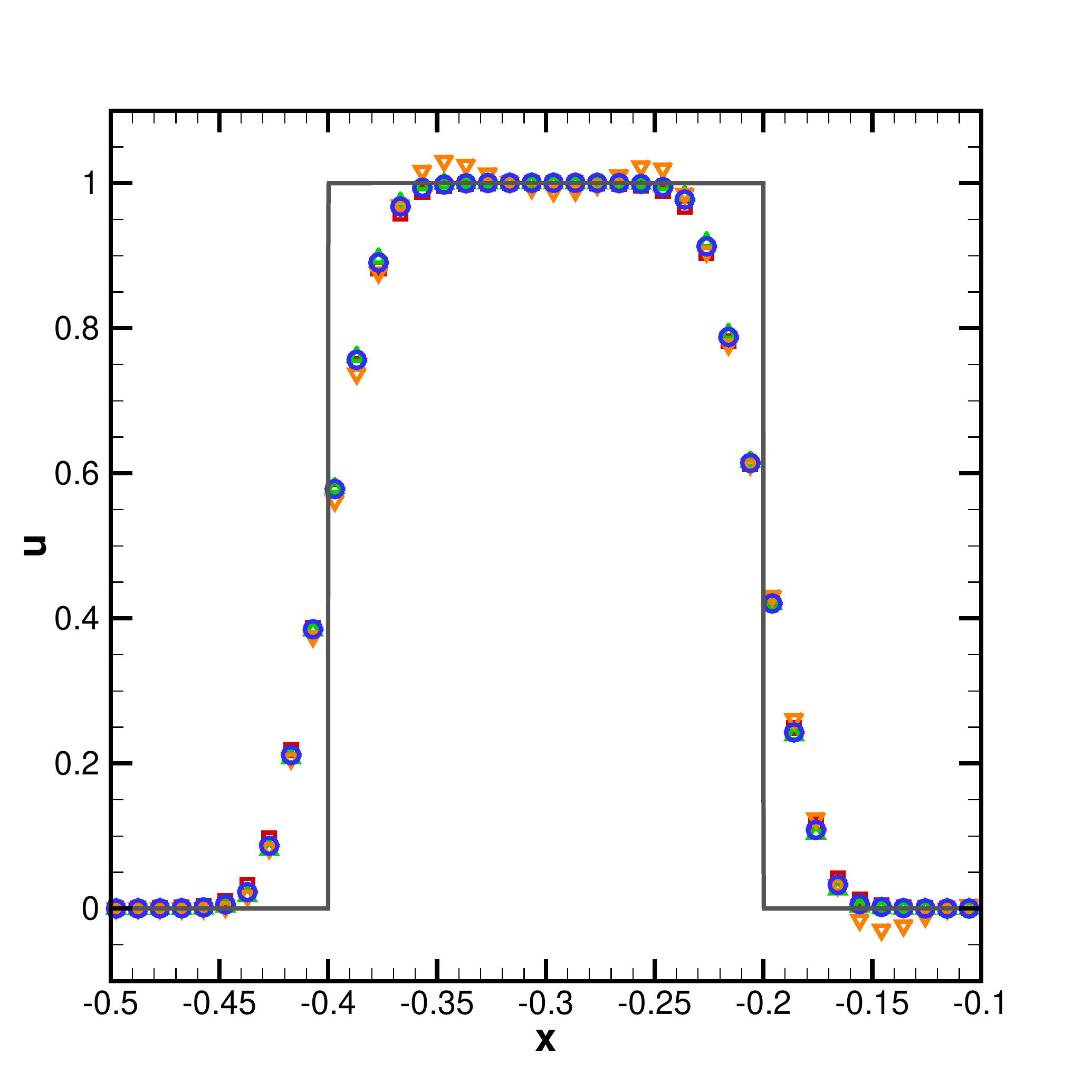}}\label{fig2:b}
  \subfigure[The triangle wave]{
  \includegraphics[width=0.45\textwidth]{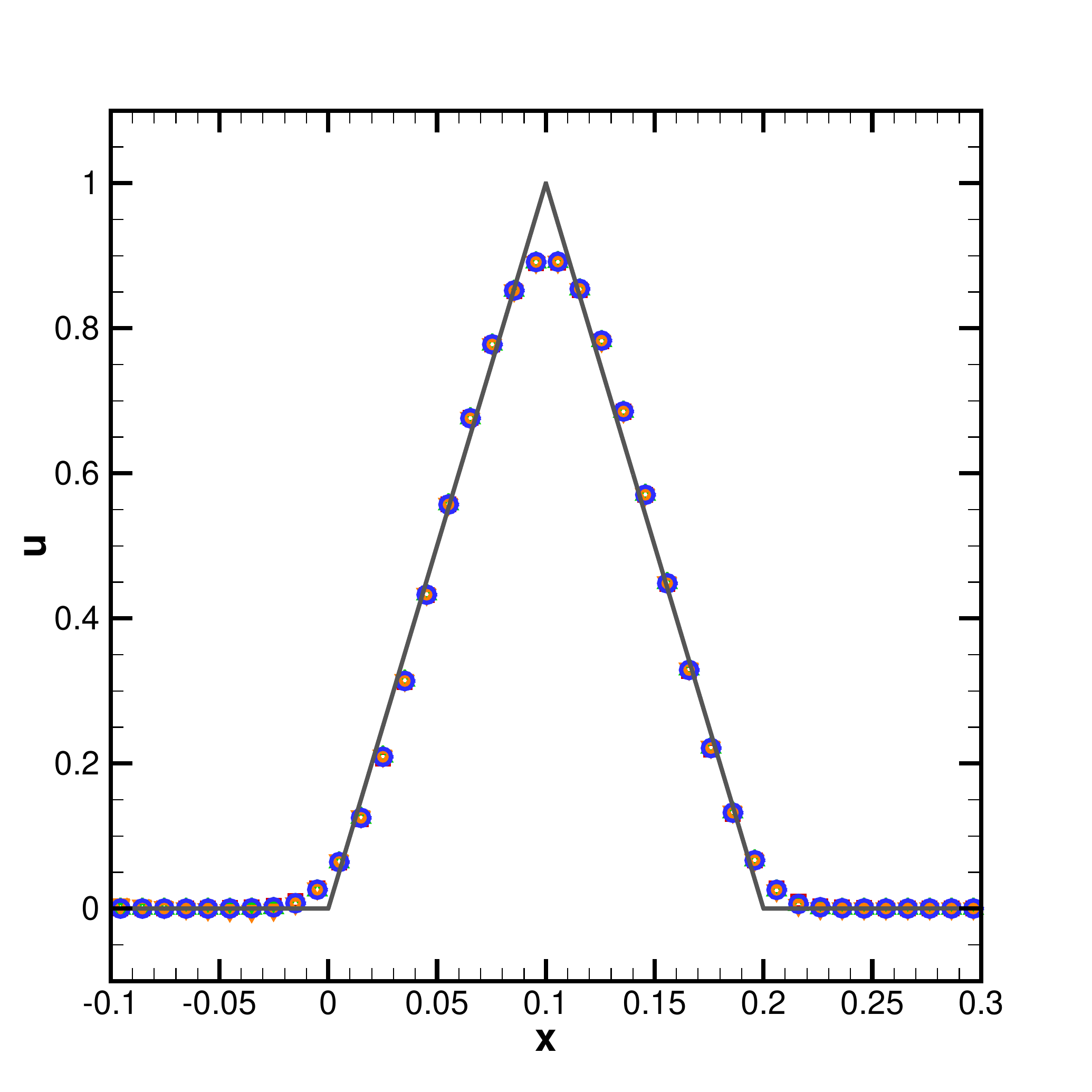}}\label{fig2:c}
  \subfigure[The ellipse wave]{
  \includegraphics[width=0.45\textwidth]{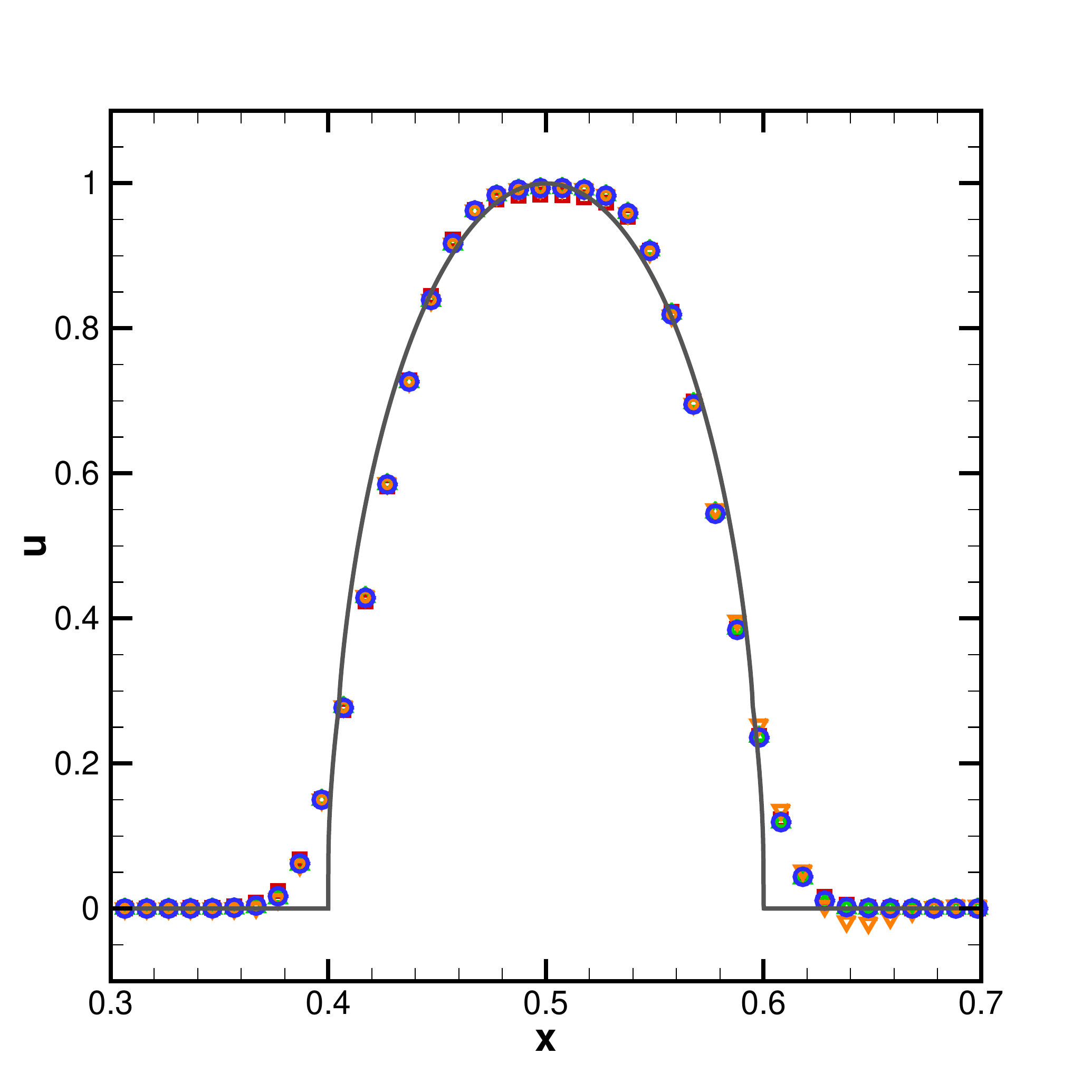}}\label{fig2:d}
  \caption{Results of the linear advection equation with the initial condition \eqref{eq:4.6} at t=6, N=200}
  \label{fig:2}
  \end{center}
\end{figure}

The initial condition of the second case is:
\begin{align}\label{eq:4.6}
& u_0(x)=
\begin{cases}
\frac{1}{6}(G(x,\beta,z-\delta)+G(x,\beta,z+\delta)+4G(x,\beta,z)), &  -0.8 \leqslant x < -0.6 \\
1, &  -0.4 \leqslant x < -0.2 \\
1-|10(x-0.1)|, &  0 \leqslant x < 0.2 \\
\frac{1}{6}(F(x,\alpha,a-\delta)+F(x,\alpha,a+\delta)+4F(x,\alpha,a)),& 0.4 \leqslant x < 0.6\\
0, & otherwise
\end{cases}&
\end{align}
where
\begin{align}
\nonumber &G(x,\beta,z)=e^{-\beta (x-z)^2}, F(x,\alpha,a)=\sqrt{max(1-\alpha^{2}(x-a)^2,0)}, \\
\nonumber &a=0.5, z=-0.7, \delta=0.005, \alpha=10, \beta=log2/36\delta^2.
\end{align}
The solution of Eq.\eqref{eq:4.6} contains a smooth but narrow combination of Gaussians, a square wave, a sharp triangle wave, and a half ellipse \cite{jiang1996efficient}. As this test case is a combination of both smooth and non-smooth functions, it has been widely used to test the discontinuity capturing capability of a scheme. Numerical results of different schemes are given in Fig.\ref{fig:2}. It can be observed that  TENO5-A produces oscillations for the square wave while the others not. the presented method well preserves the ENO property for the discontinuities and maintains low dissipation for the smooth waves.

\subsection{One dimensional Euler equations}\label{sec4.2}
The one dimensional Euler equations are given by
\begin{align}\label{eq:4.7}
U_t+F(U)_x=0 
\end{align}
where $U=(\rho,\rho u,e)^T$ and $F(U)=(\rho u,\rho u^2+p, u(e+p))^T$. Here $\rho$ is the density, $u$ is the velocity, $e$ is the total energy, $p$ is the pressure, and for ideal gas $e=\frac{p}{\gamma-1}+\frac{1}{2}\rho u^2$, $\gamma=1.4$ is the ratio of specific heat.

Four typical examples containing strong discontinuities are considered here. The first example is the Sod problem. The initial conditions are:
\begin{align}\label{eq:4.8} 
 &(\rho,u,p)=
\begin{cases}
(1,0,1)& x \leqslant 0 \\
(0.125,0,0.1)& x>0
\end{cases}&
\end{align}
with zero gradient boundary conditions applied at $x=\pm 0.5$.

The second problem is the Lax problem, the initial conditions are given by:
\begin{align}\label{eq:4.9} 
 &(\rho,u,p)=
\begin{cases}
(0.445,0.698,3.528)& x\leqslant0\\
(0.5,0,0.571)& x>0
\end{cases}&
\end{align}
with zero gradient boundary conditions at $x=\pm5$.

The third problem is the Shu-Osher problem. It describes the interaction of a Mach 3 shock with a density wave. The initial conditions are given by:
\begin{align}\label{eq:4.10} 
 &(\rho,u,p)=
\begin{cases}
(\frac{27}{7},\frac{4\sqrt{35}}{9},\frac{31}{3})& x<-4\\
(1+\frac{1}{5}sin5x,0,1)& x\geqslant -4
\end{cases}&
\end{align}
Zero gradient boundary conditions are applied at $x=\pm5$.

The last problem is the two interacting blast waves case. The initial conditions are given by:
\begin{align}\label{eq:4.11} 
 &(\rho,u,p)=
\begin{cases}
(1,0,1000)& 0\leqslant x<0.1\\
(1,0,0.01)& 0\leqslant x<0.9\\
(1,0,100)& 0.9\leqslant x\leqslant 1.
\end{cases}&
\end{align}
Reflecting boundary conditions are set at boundaries.

\begin{figure}[H]
  \begin{center}
  \subfigure[Density distribustion]{
  \includegraphics[width=0.45\textwidth]{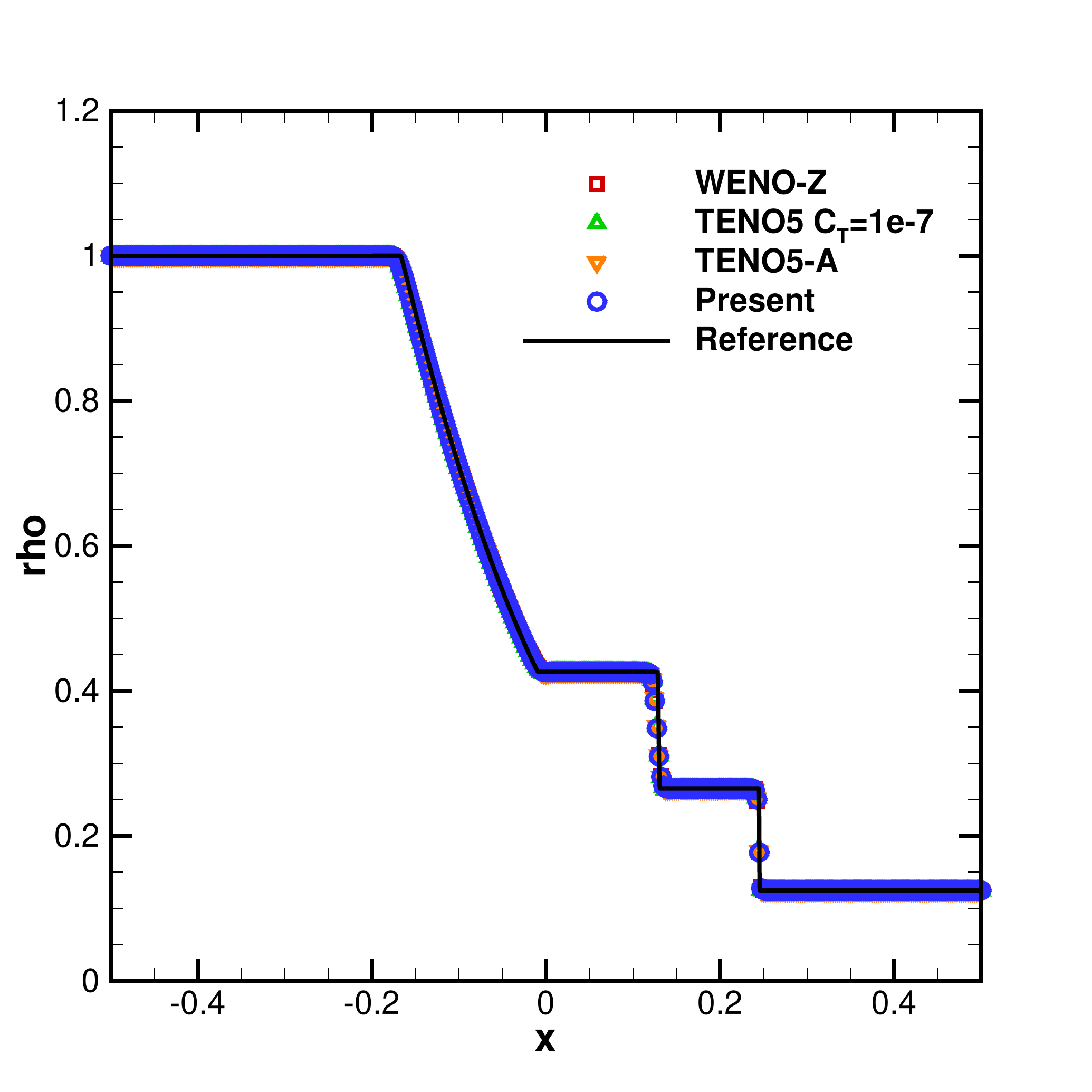}}\label{fig:sod}
  \subfigure[Zoom-in view]{
  \includegraphics[width=0.45\textwidth]{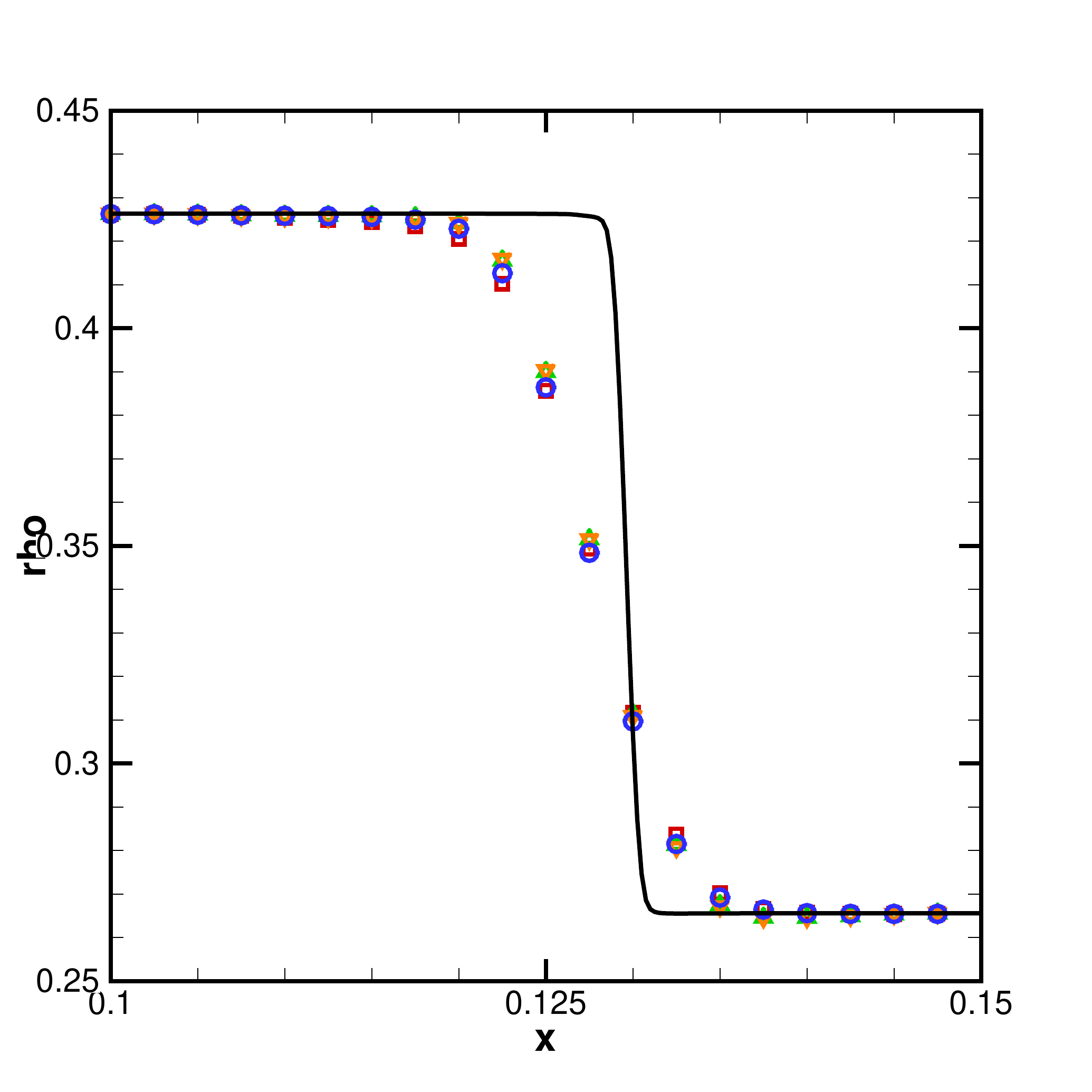}}\label{fig:lax}
  \caption{Results of different schemes for the Sod problem, t=0.14, N=400}
\label{fig:3}
\end{center}
\end{figure}

\begin{figure}[H]
  \begin{center}
  \subfigure[Density distribustion]{
  \includegraphics[width=0.45\textwidth]{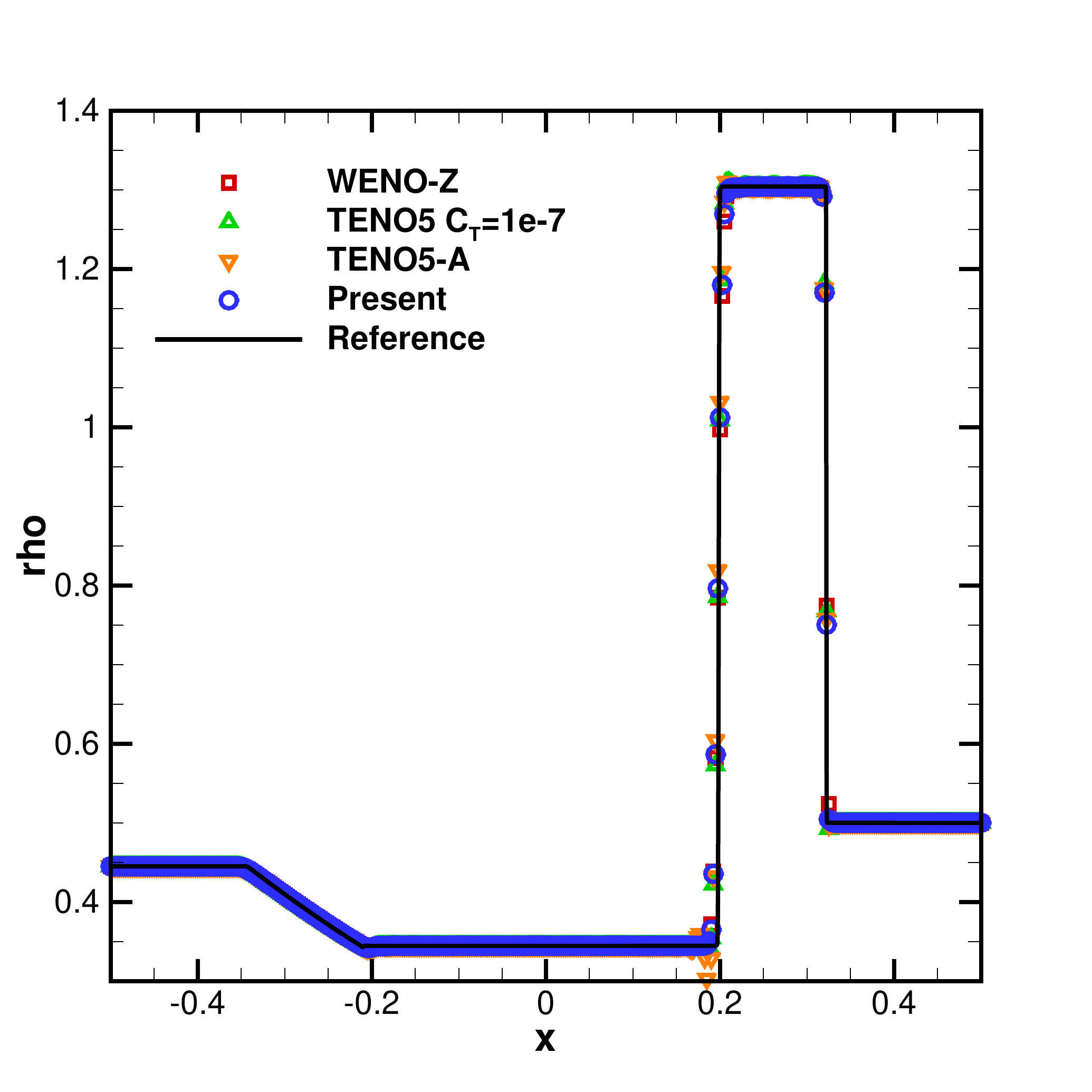}}\label{fig:sod}
  \subfigure[Zoom-in view]{
  \includegraphics[width=0.45\textwidth]{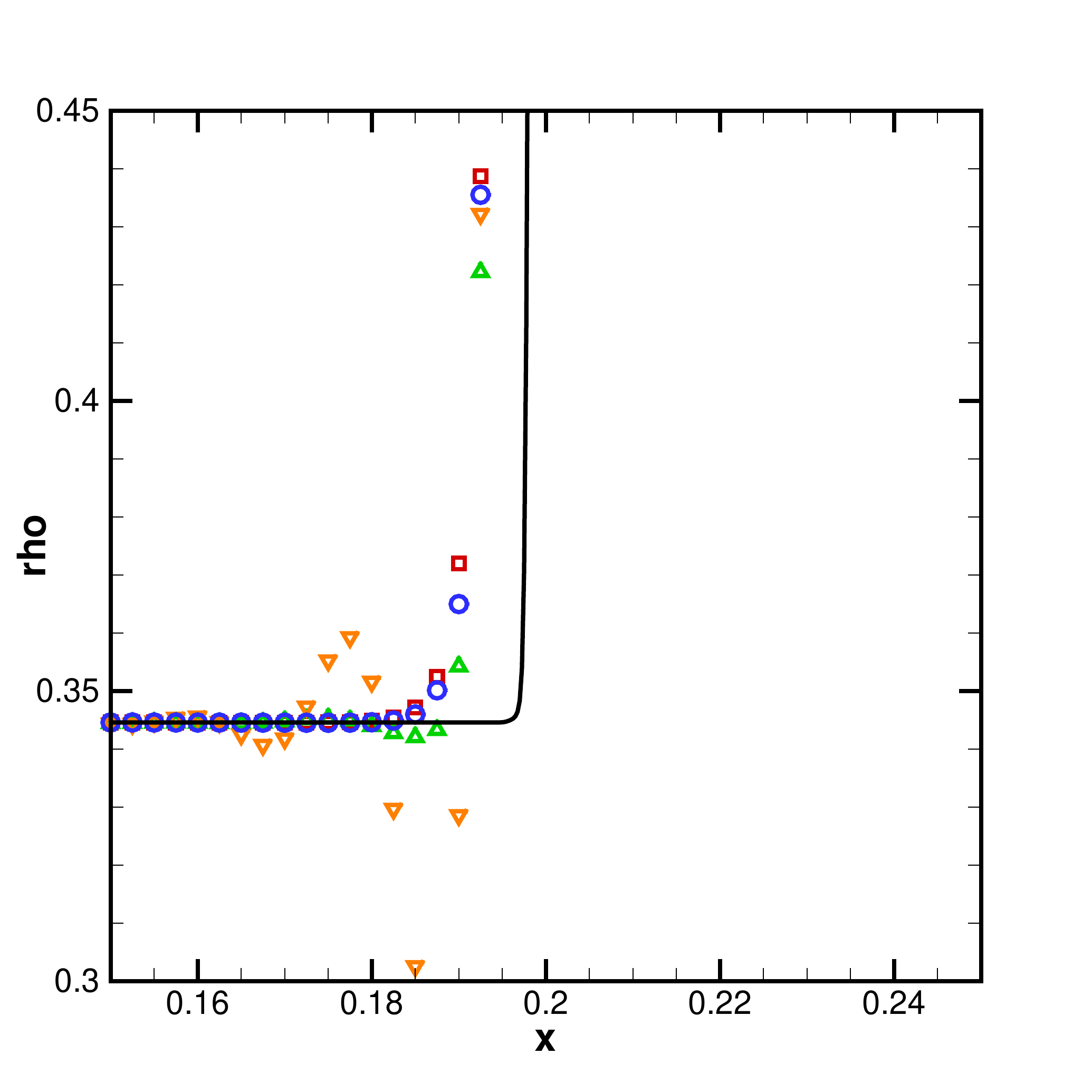}}\label{fig:lax}
  \caption{Results of different schemes for the Lax problem, t=0.13, N=400}
\label{fig:4}
\end{center}
\end{figure}

\begin{figure}[H]
  \begin{center}
  \subfigure[Density distribustion, N=200]{
  \includegraphics[width=0.45\textwidth]{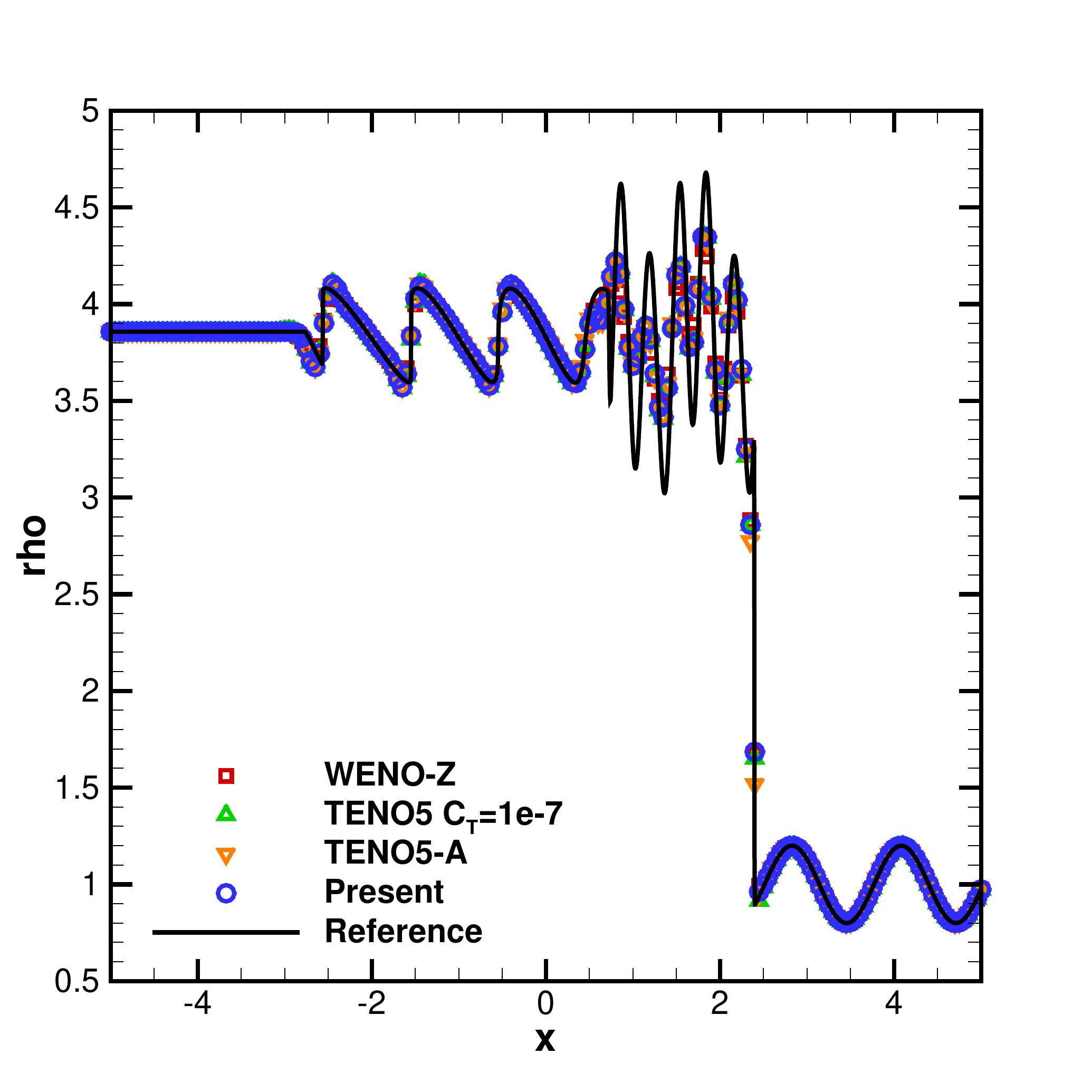}\label{fig:so200a}}
  \subfigure[Zoom-in view of (a)]{
  \includegraphics[width=0.45\textwidth]{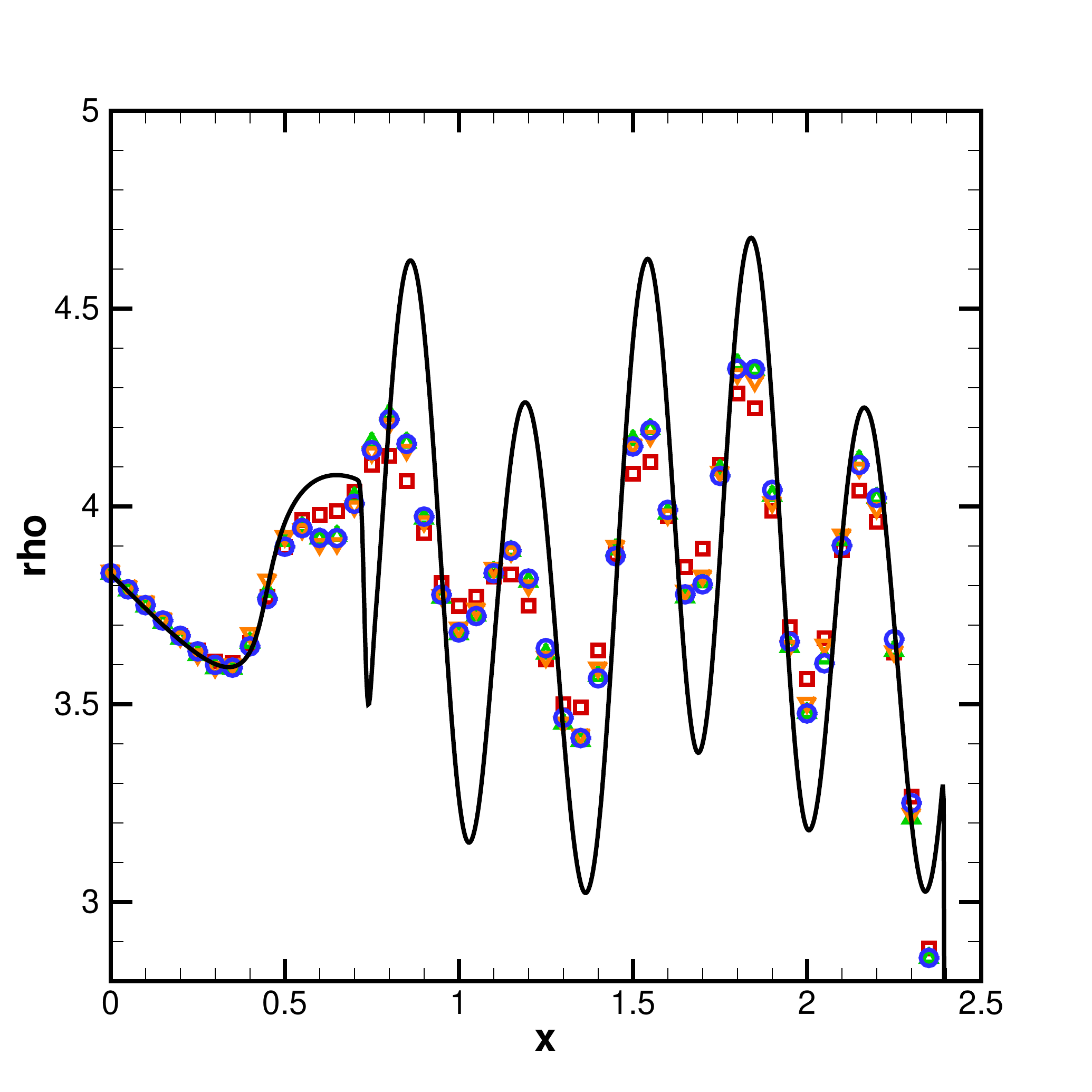}\label{fig:so200b}}
  \subfigure[Density distribustion, N=400]{
  \includegraphics[width=0.45\textwidth]{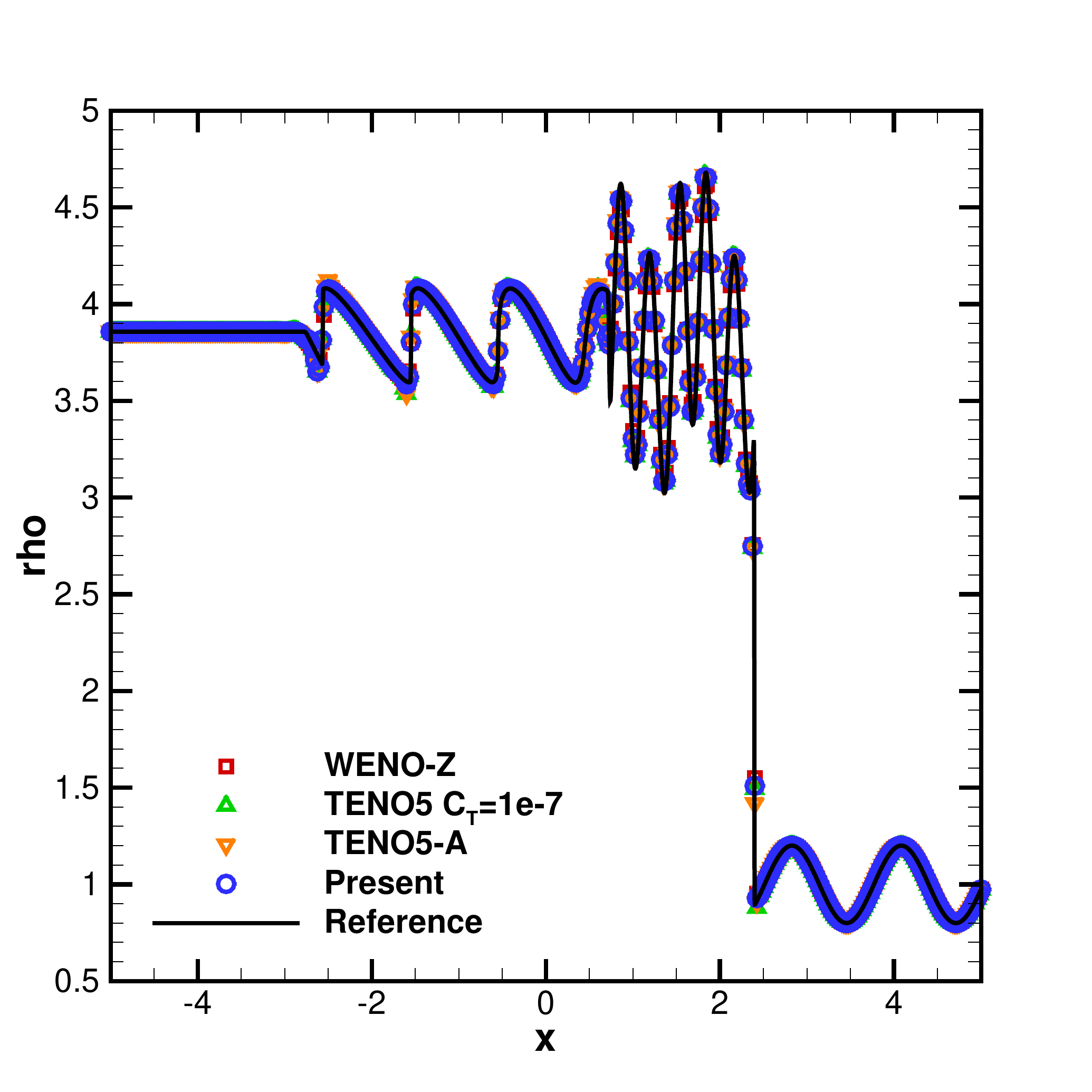}\label{fig:so400a}}
  \subfigure[Zoom-in view of (c)]{
  \includegraphics[width=0.45\textwidth]{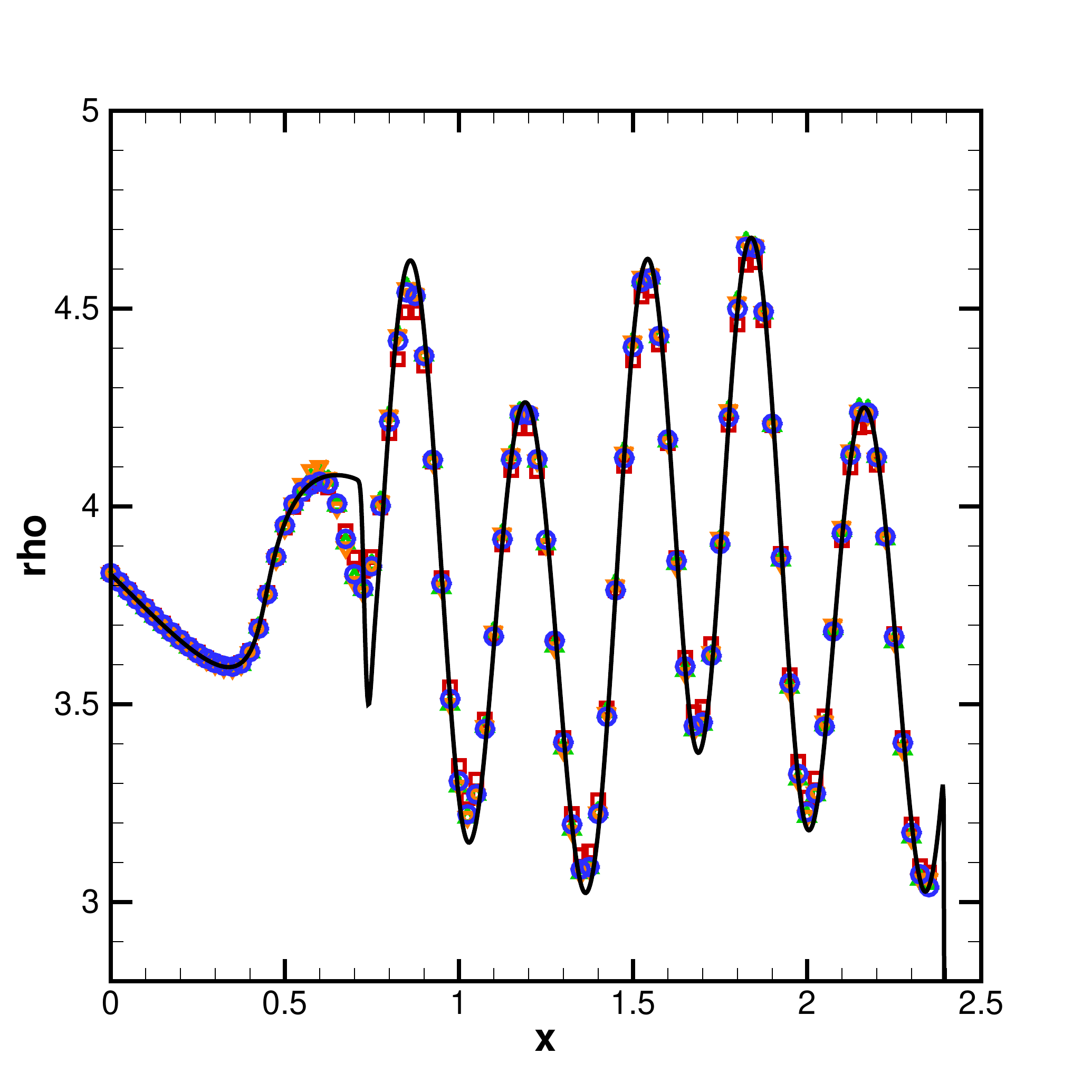}\label{fig:so400b}}
  \caption{Results of different schemes for the Shu-Osher problem, t=1.8}\label{fig:5}
\end{center}
\end{figure}

\begin{figure}[H]
  \begin{center}
  \subfigure[Density distribustion]{
  \includegraphics[width=0.45\textwidth]{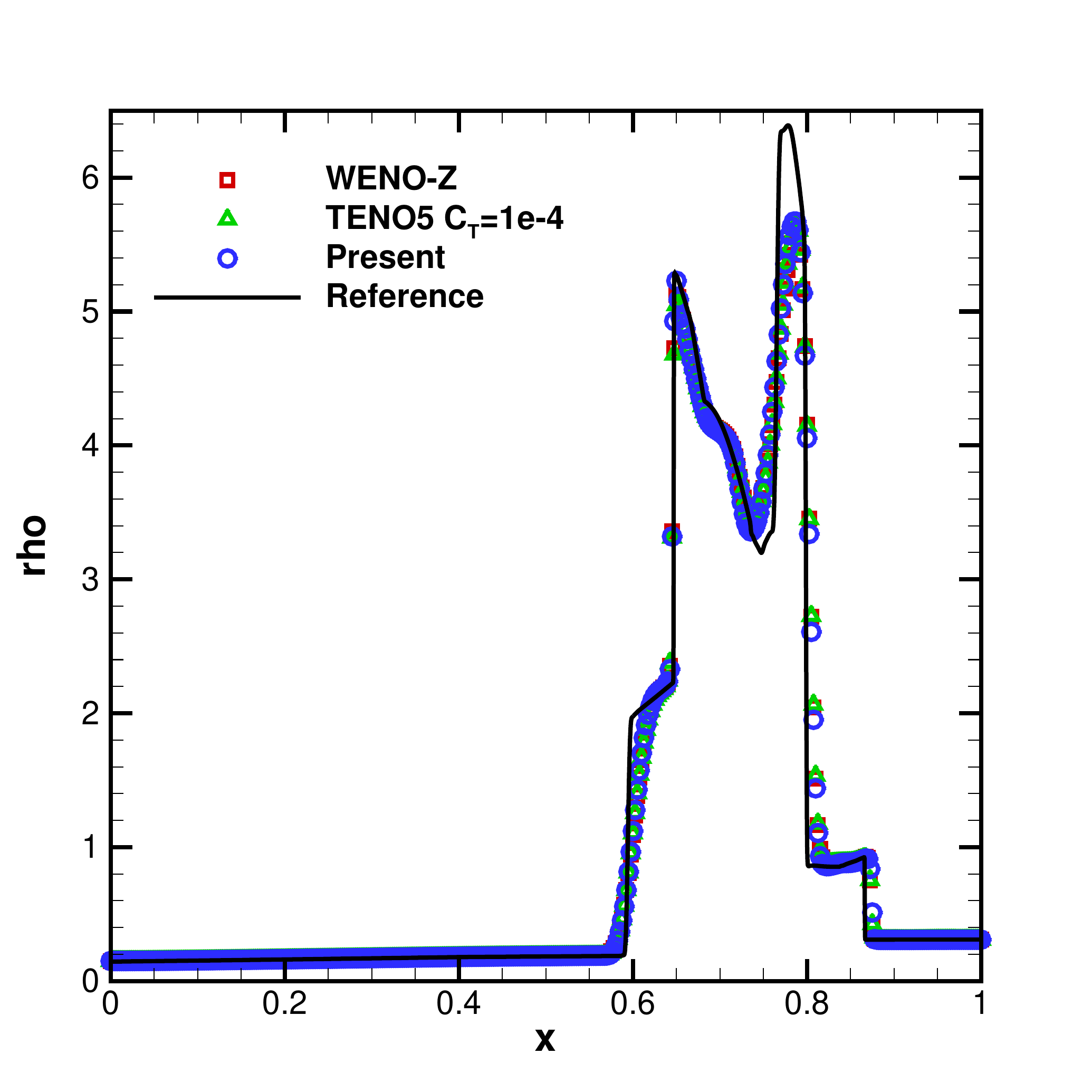}\label{fig:tbwa}}
  \subfigure[Zoom-in view]{
  \includegraphics[width=0.45\textwidth]{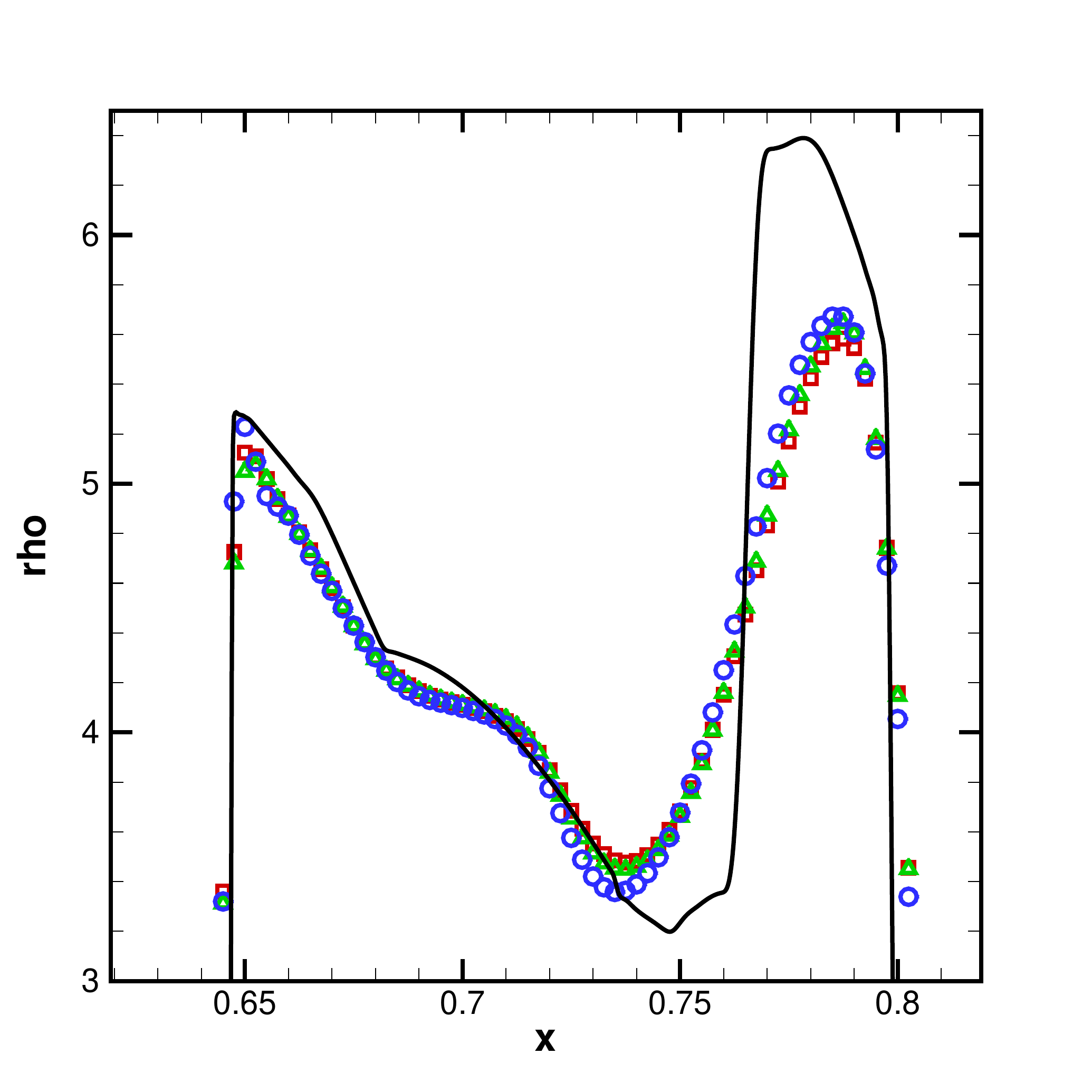}\label{fig:tbwb}}
  \caption{Results of different schemes for the two interacting blast waves problem, t=0.038, N=400}
\label{fig:6}
\end{center}
\end{figure}
Fig.\ref{fig:3} to Fig.\ref{fig:6} illustrate results of the above cases. Reference results are obtained by the WENO-Z scheme with 2000 points. 

For the sod problem, as shown in Fig.\ref{fig:3}, each scheme resolves the discontinuities well. When the discontinuity becomes stronger, as being shown in Fig.\ref{fig:4}, both of the TENO5 scheme and the TENO5-A method produce oscillations near the contact wave. the presented method maintains the ENO property. 

For the result of the Shu-Osher problem (Fig.\ref{fig:5}), on the coarse grid (N=200), the TENO methods give better resolved short waves than the WENO-Z method. On the fine mesh (N=400), overshoot can be observed from the results of TENO5-A (Fig.\ref{fig:so400b}), while the presented method still keeps the ENO property as well as low dissipation for the short waves. 

The two interacting blast waves problem is a more severe case with very strong discontinuities for high order schemes. With the given parameters, the TENO5-A method blows up and therefore no result is given in Fig.\ref{fig:6}. $C_T$ is adjusted to $10^{-4}$ for TENO5 to obtain a stable result. the presented method requires no parameter tuning. From Fig.\ref{fig:tbwb}, it can be seen that the presented method obtains better resolved structures than the others.

\subsection{Two dimensional Euler equations}\label{sec4.3}

\subsubsection{Rayleigh–Taylor instability}\label{sec4.3.2}
The two-dimensional Rayleigh-Taylor instability problem is often used to assess the dissipation property of a high-order scheme. It describes the interface instability between fluids with different densities when acceleration is directed from the heavy fluid to the light one. The acceleration effect is introduced by adding $\rho$ and $\rho{v}$ to the $y$-momentum and the energy equations, respectively. The initial conditions are:
\begin{align}\label{eq:4.12}
&(\rho,u,v,p)=
\begin{cases}
(2,0,-0.025 \alpha cos(8\pi{x}),2y+1),& 0\le{y}<1/2, \\
(1,0,-0.025 \alpha cos(8\pi{x}),y+3/2),& 1/2\le{y}<1,
\end{cases}&
\end{align}
where $\alpha=\sqrt{\gamma{p}/\rho}$ is the speed of sound with $\gamma=5/3$. The computational domain is $[0,0.25]\times{[0,1]}$. The left and right boundaries are set with reflective boundary conditions, and the top and bottom boundaries are set as $(\rho,u,v,p)=(1,0,0,2.5)$ and $(\rho,u,v,p)=(2,0,0,1)$. The solution is integrated to $t=1.95$.
\begin{figure}[H]
  \begin{center}
  \subfigure[WENO-Z]{
  \includegraphics[width=0.23\textwidth]{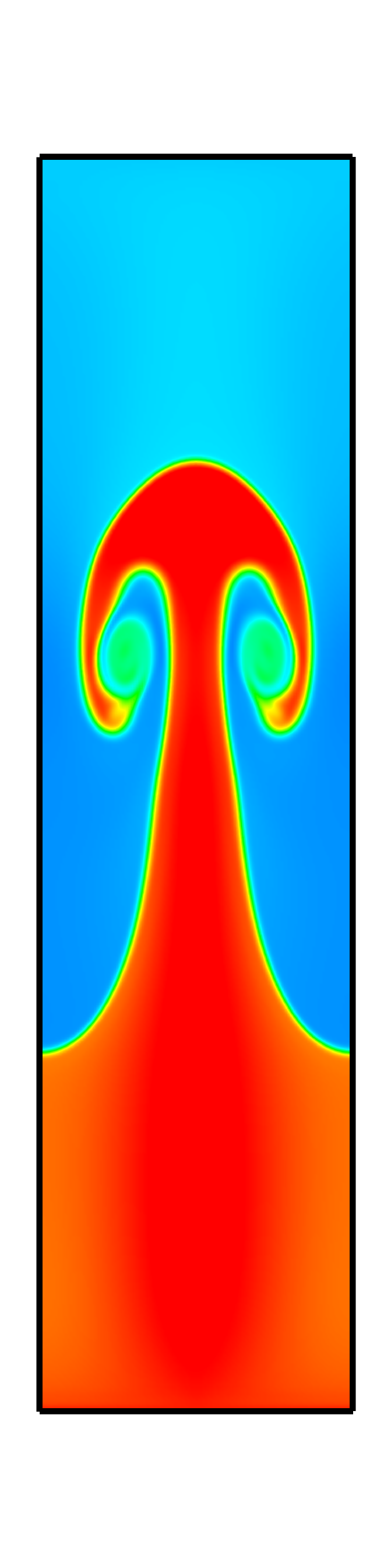}}
  \subfigure[TENO5]{
  \includegraphics[width=0.23\textwidth]{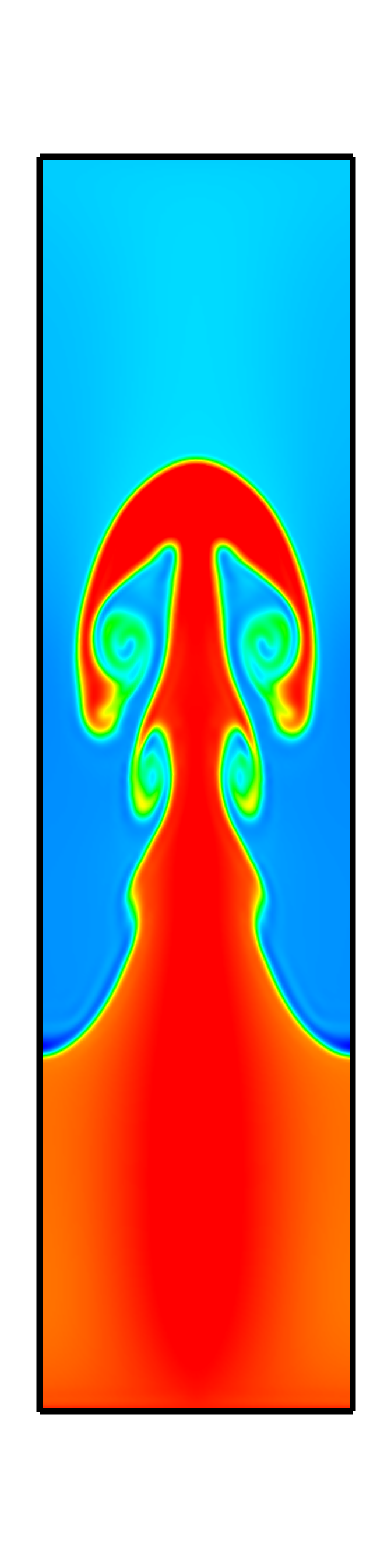}}
  \subfigure[TENO5-A]{
  \includegraphics[width=0.23\textwidth]{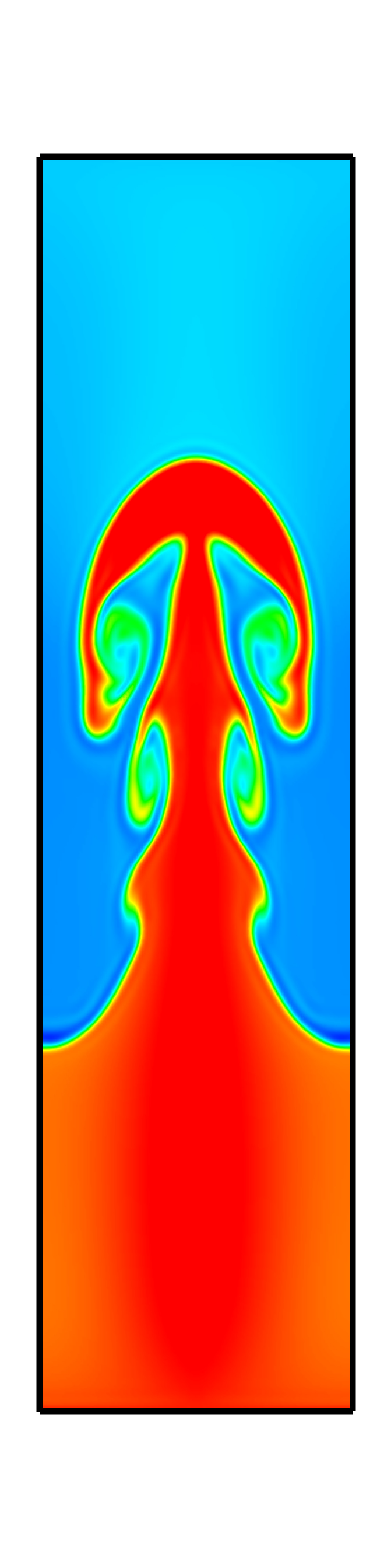}}
  \subfigure[Present]{
  \includegraphics[width=0.23\textwidth]{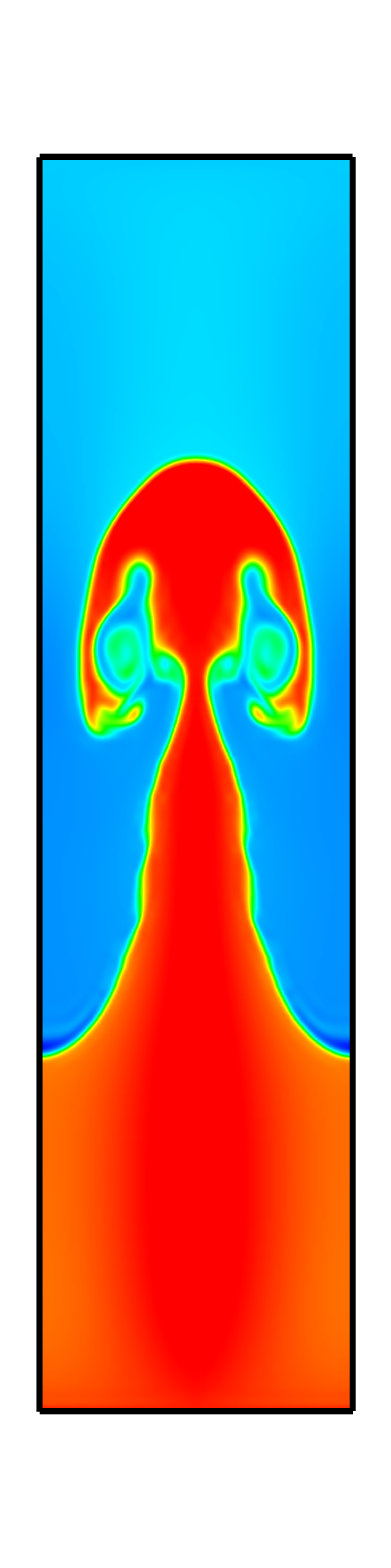}}
  \caption{Density contours of different schemes for the Rayleigh–Taylor instability problem at t=1.95, $[N_x \times N_y]=[128 \times 256]$.}
\label{fig:7}
\end{center}
\end{figure}

\begin{figure}[H]
  \begin{center}
  \subfigure[WENO-Z]{
  \includegraphics[width=0.23\textwidth]{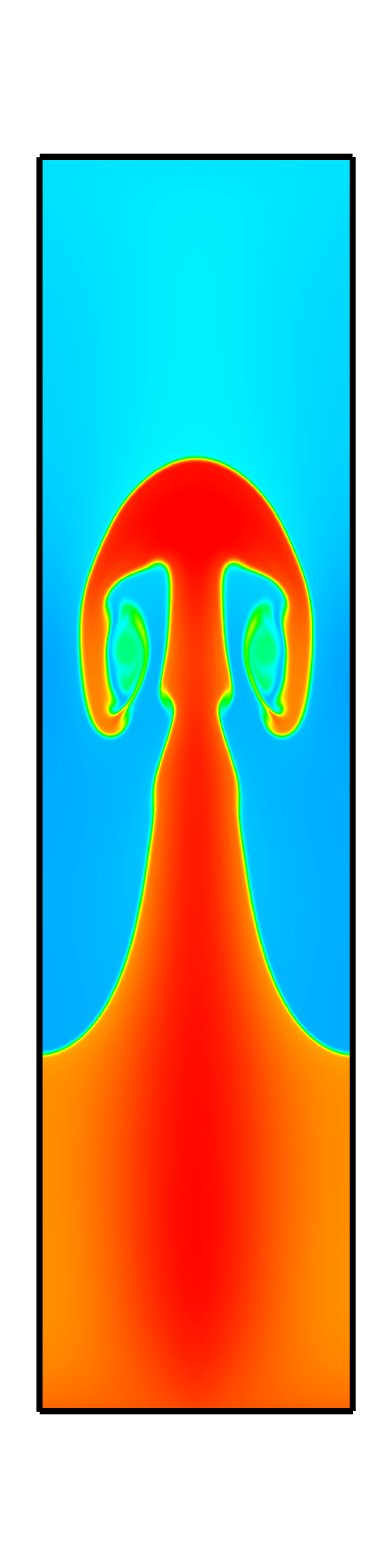}}
  \subfigure[TENO5]{
  \includegraphics[width=0.23\textwidth]{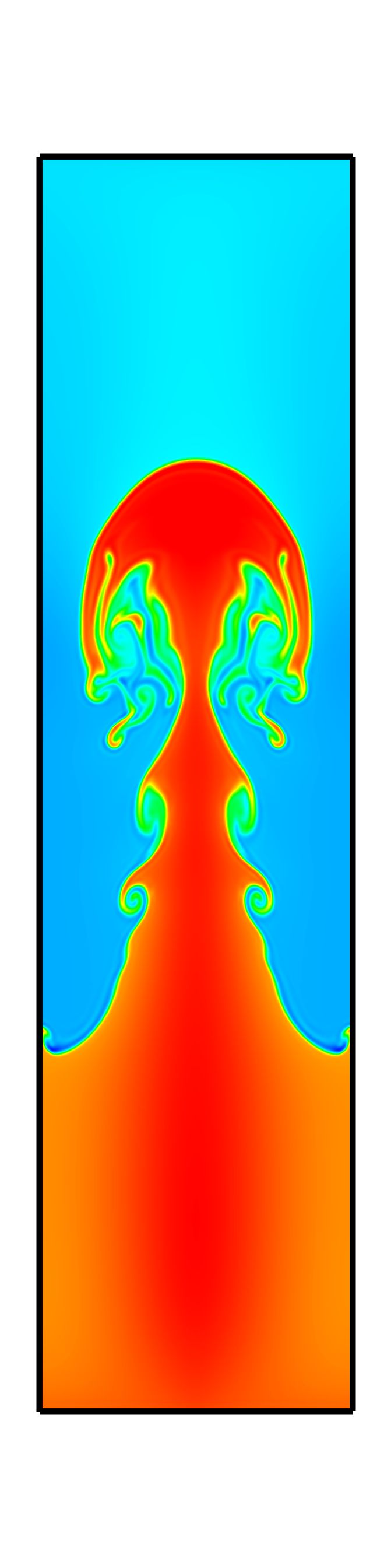}}
  \subfigure[TENO5-A]{
  \includegraphics[width=0.23\textwidth]{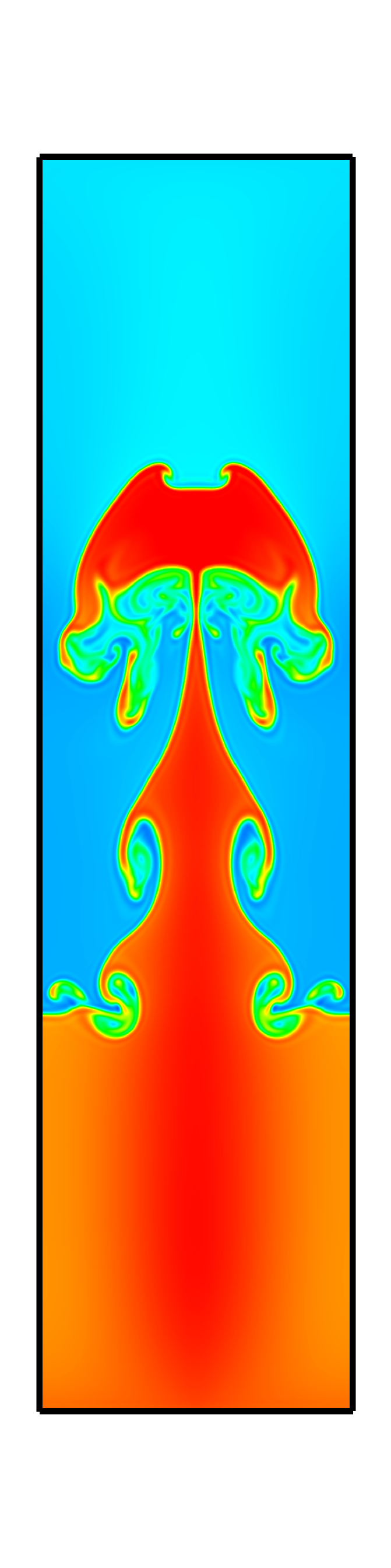}}
  \subfigure[Present]{
  \includegraphics[width=0.23\textwidth]{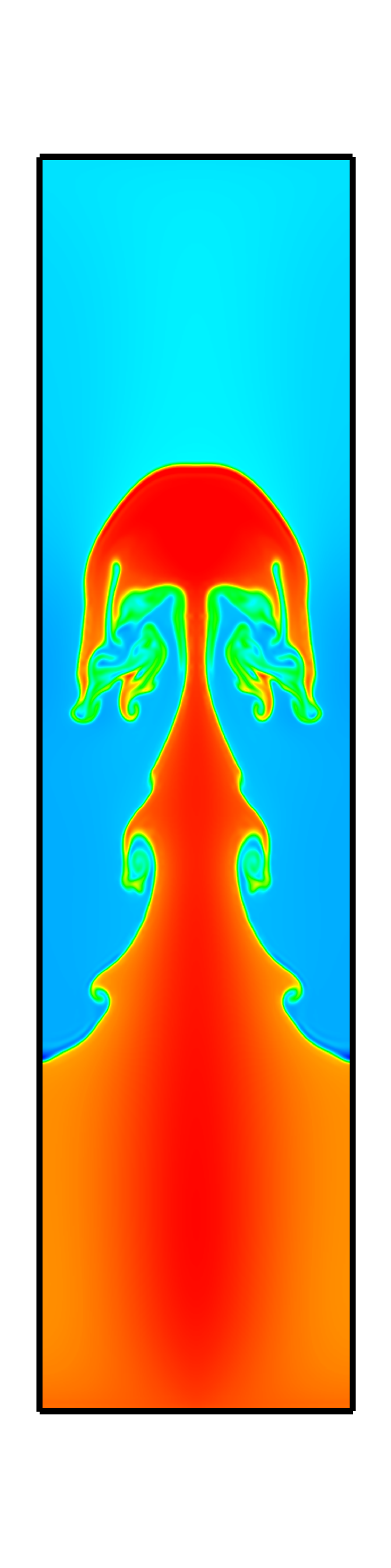}}
  \caption{Density contours of different schemes for the Rayleigh–Taylor instability problem at t=1.95, $[N_x \times N_y]=[256 \times 512]$}
\label{fig:8}
\end{center}
\end{figure}

Solutions on two sets of meshes are illustrated in Fig.\ref{fig:7} and Fig.\ref{fig:8}. On the coarse mesh ($[N_x \times N_y]=[128 \times 256]$), the density contours in Fig.\ref{fig:7} show that the TENO schemes resolve more structures than the WENO-Z scheme. Although TENO5 and TENO5-A give richer small structures, they also produce obvious oscillations in vicinity of the interface. The present scheme is able to suppress such oscillations. On the fine mesh ($[N_x \times N_y]=[256 \times 512]$), small structures are also better resolved by the TENO schemes. As with on the coarse mesh, TENO5 and TENO5-A still produce oscillations near the interface, the presented method maintains high resolution for the small structures and reduces numerical oscillations. 

\subsubsection{Riemann problems}\label{sec4.3.1}
Two 2D Riemann problems are considered in this section. 

The first case corresponds to configuration 3 in \cite{Kurganov2002} with the initial conditions
\begin{align}\label{eq:4.14} 
&(\rho,u,v,p)=
\begin{cases}
(0.138,\ 1.206,\ 1.206,\ 0.029) & x \le 0.5 , y \le 0.5, \\
(0.5323,\ 1.206,\ 0.0,\ 0.3) & x \le 0.5 , y > 0.5, \\
(0.5323,\ 0.0,\ 1.206,\ 0.3) & x > 0.5 , y \le 0.5, \\
(1.5,\ 0.0,\ 0.0,\ 1.5) & x > 0.5 , y > 0.5.
\end{cases}&
\end{align}
Solutions are integrated to $t=0.3$. A uniform grid of $(1024 \times 1024)$ is used.

\begin{figure}[H]
  \begin{center}
  \subfigure[WENO-Z]{
  \includegraphics[width=0.45\textwidth]{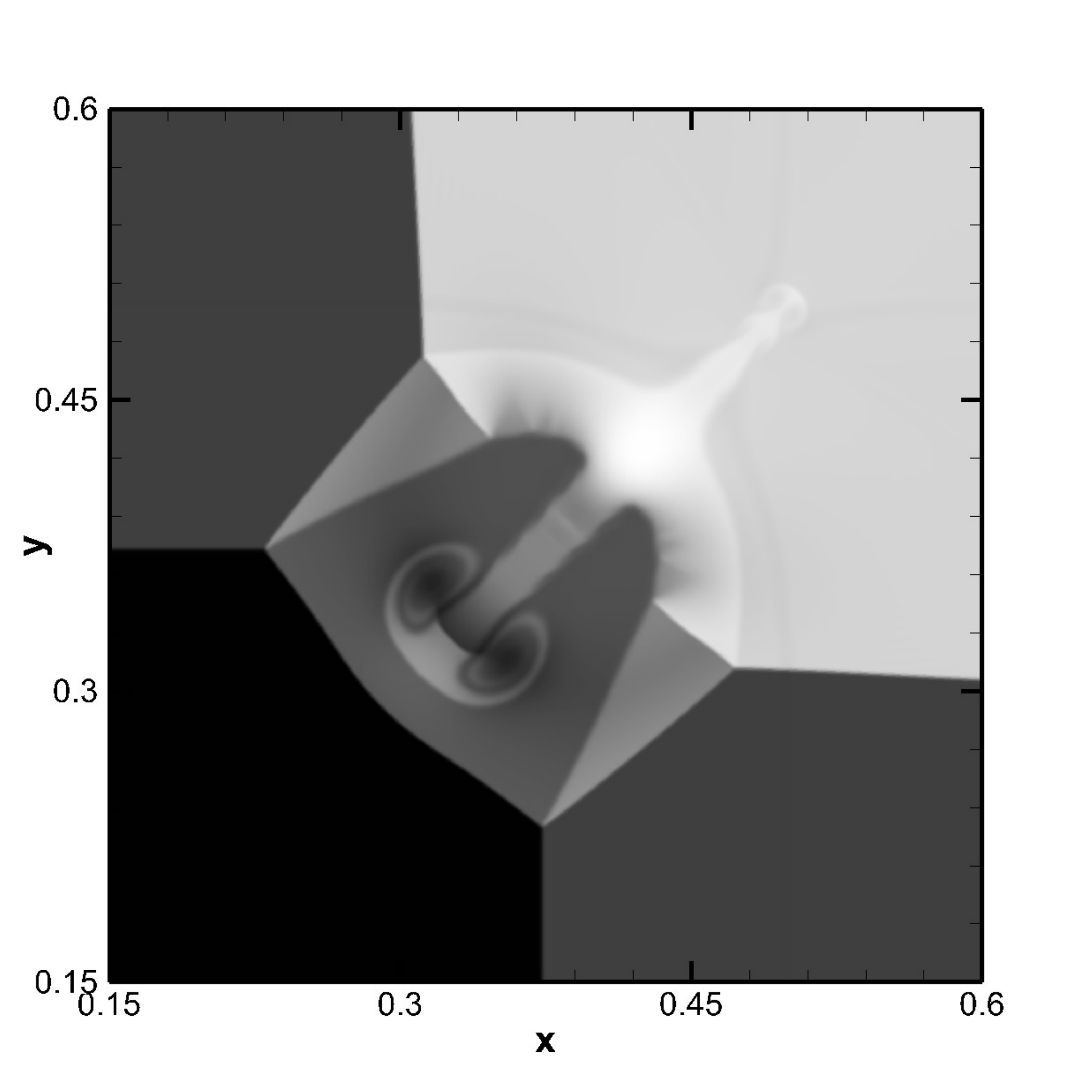}}
  \subfigure[TENO5]{
  \includegraphics[width=0.45\textwidth]{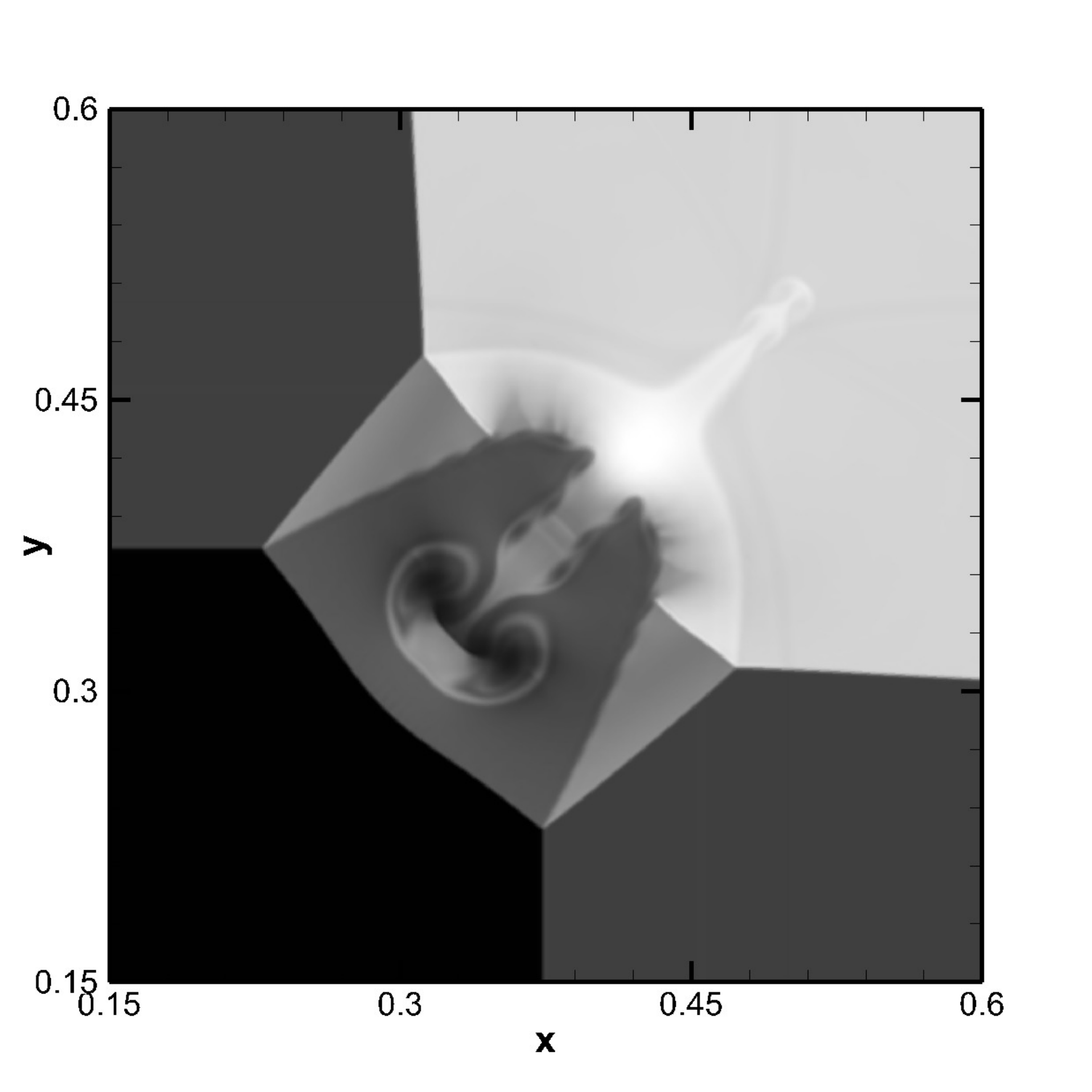}}
  \subfigure[TENO5-A]{
  \includegraphics[width=0.45\textwidth]{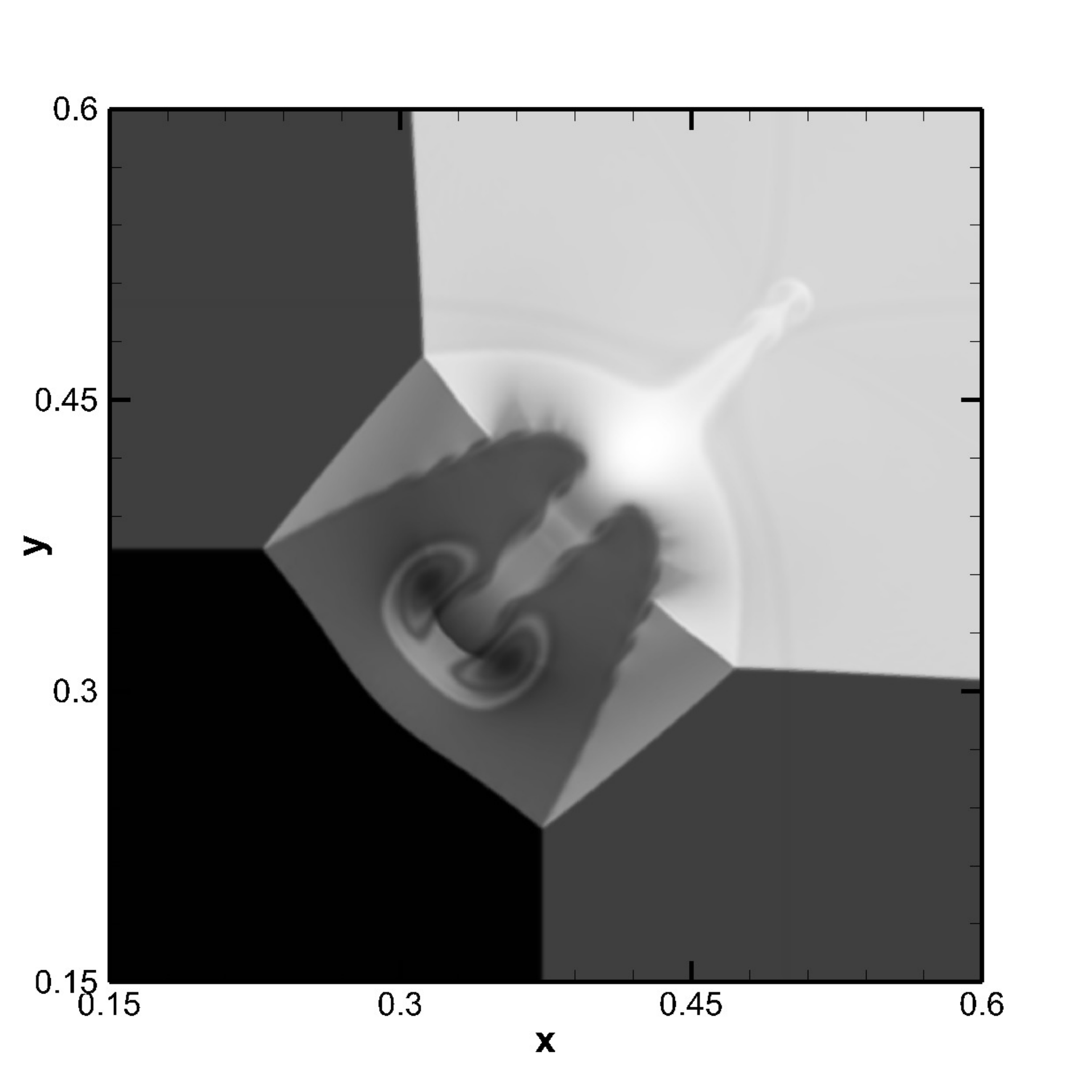}}
  \subfigure[Present]{
  \includegraphics[width=0.45\textwidth]{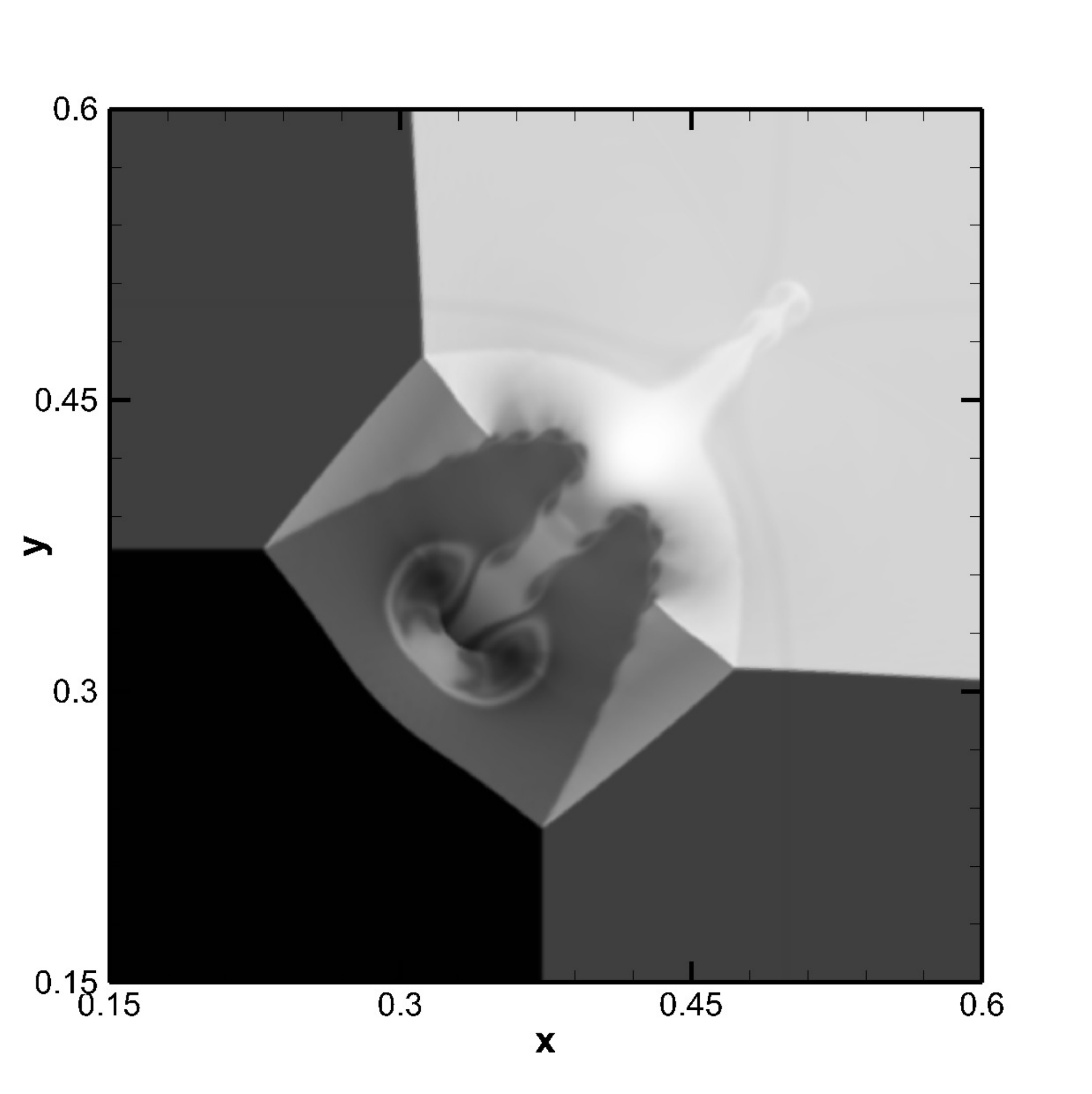}}
  \caption{Density contours for 2D Riemann problem with initial conditions Eq.\eqref{eq:4.14}}
  \label{fig:10}
\end{center}
\end{figure}

For this case, the parameters of the TENO5-A scheme are adjusted as:
$$
a_1 = 10.5, \quad a_2 = 5.5, \quad C_r=0.3, \quad \xi = 10^{-3},
$$
otherwise the computation will blow up. Density contours are shown in Fig.\ref{fig:10}. It can be seen that the TENO schemes present much richer KH instability structures than the WENO-Z scheme. The result of TENO5-LAD is less dissipative than the other two TENO schemes.

The second case corresponds to configuration 12 in \cite{Kurganov2002} with the initial conditions
\begin{align}\label{eq:4.13}
&(\rho,u,v,p)=
\begin{cases}
(0.8,\ 0,\ 0,\ 1.0)        & x \le 0.5 , y \le 0.5, \\
(1.0,\ 0.7276,\ 0,\ 1.0)   & x \le 0.5 , y > 0.5, \\
(1.0,\ 0.0,\ 0.7276,\ 1.0) & x > 0.5 , y \le 0.5, \\
(0.5313,\ 0.0,\ 0.0,\ 0.4) & x > 0.5 , y > 0.5.
\end{cases}&
\end{align}
Solutions are integrated to $t=0.25$. As the small structures induced by the KH instability for this case require even lower dissipation and smaller grid size \cite{Deng2019}, two sets of grids of $(1024 \times 1024)$ and $(2048 \times 2048)$ are used to illustrate the advantage of the presented method. 

\begin{figure}[H]
  \begin{center}
  \subfigure[WENO-Z]{
  \includegraphics[width=0.45\textwidth]{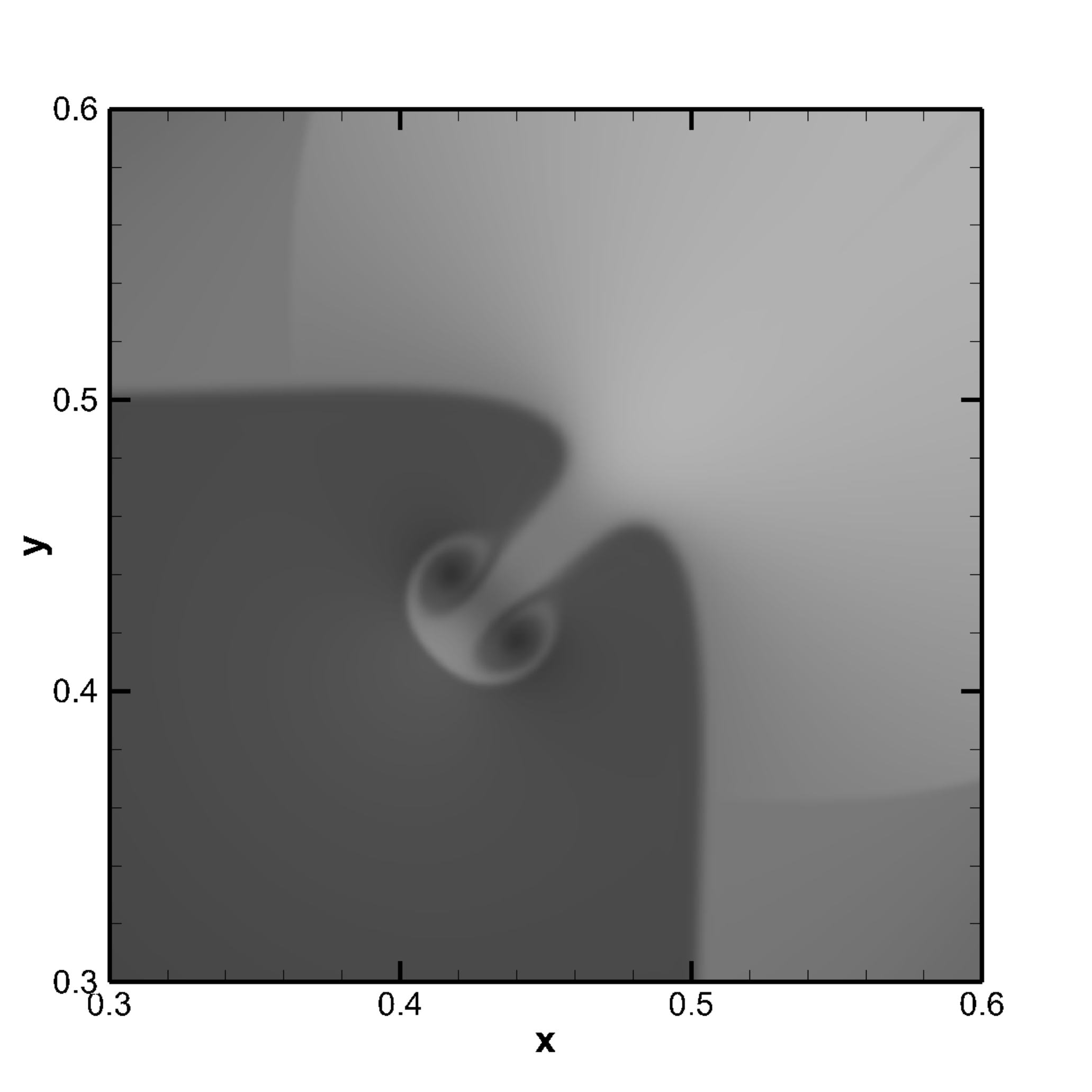}}
  \subfigure[TENO5]{
  \includegraphics[width=0.45\textwidth]{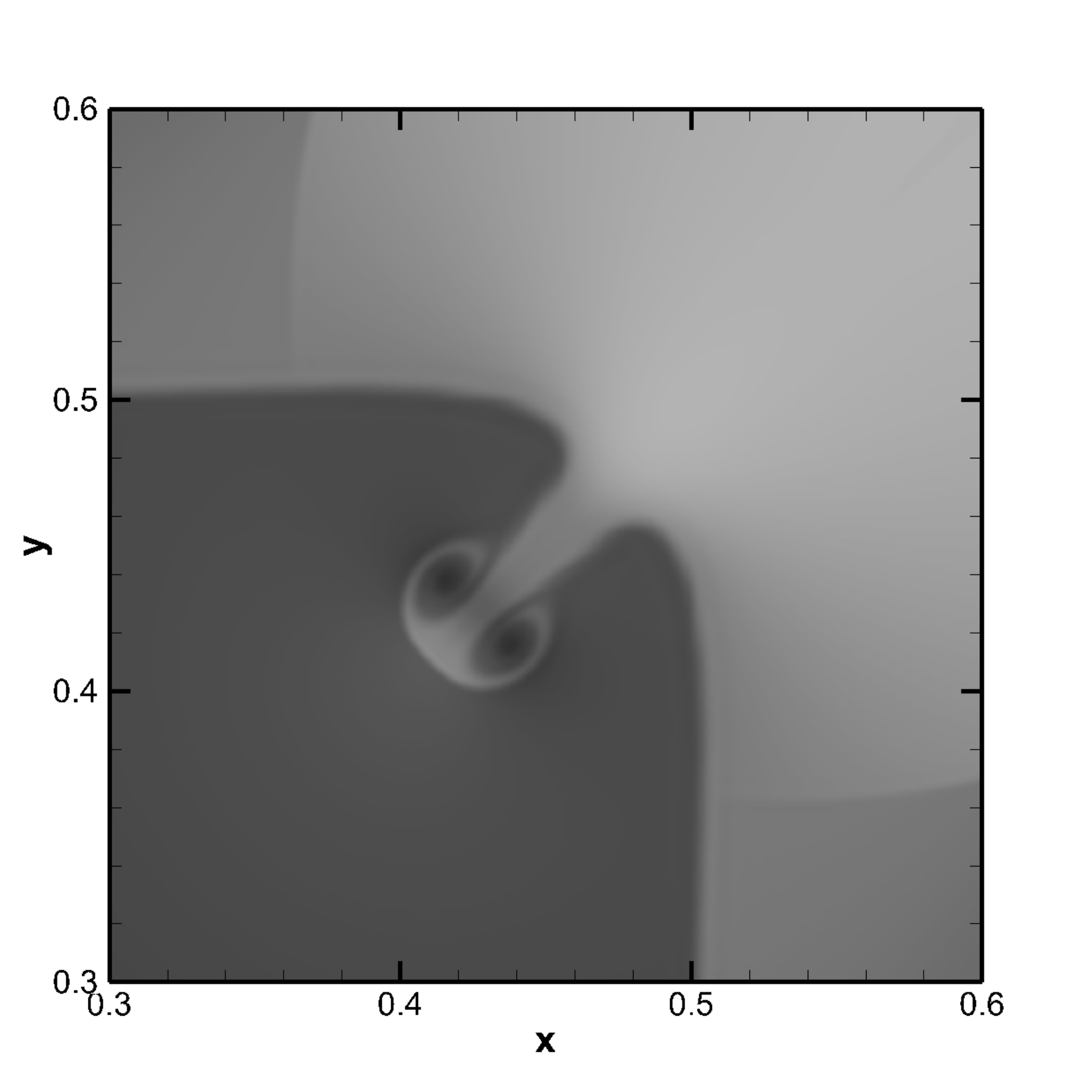}}
  \subfigure[TENO5-A]{
  \includegraphics[width=0.45\textwidth]{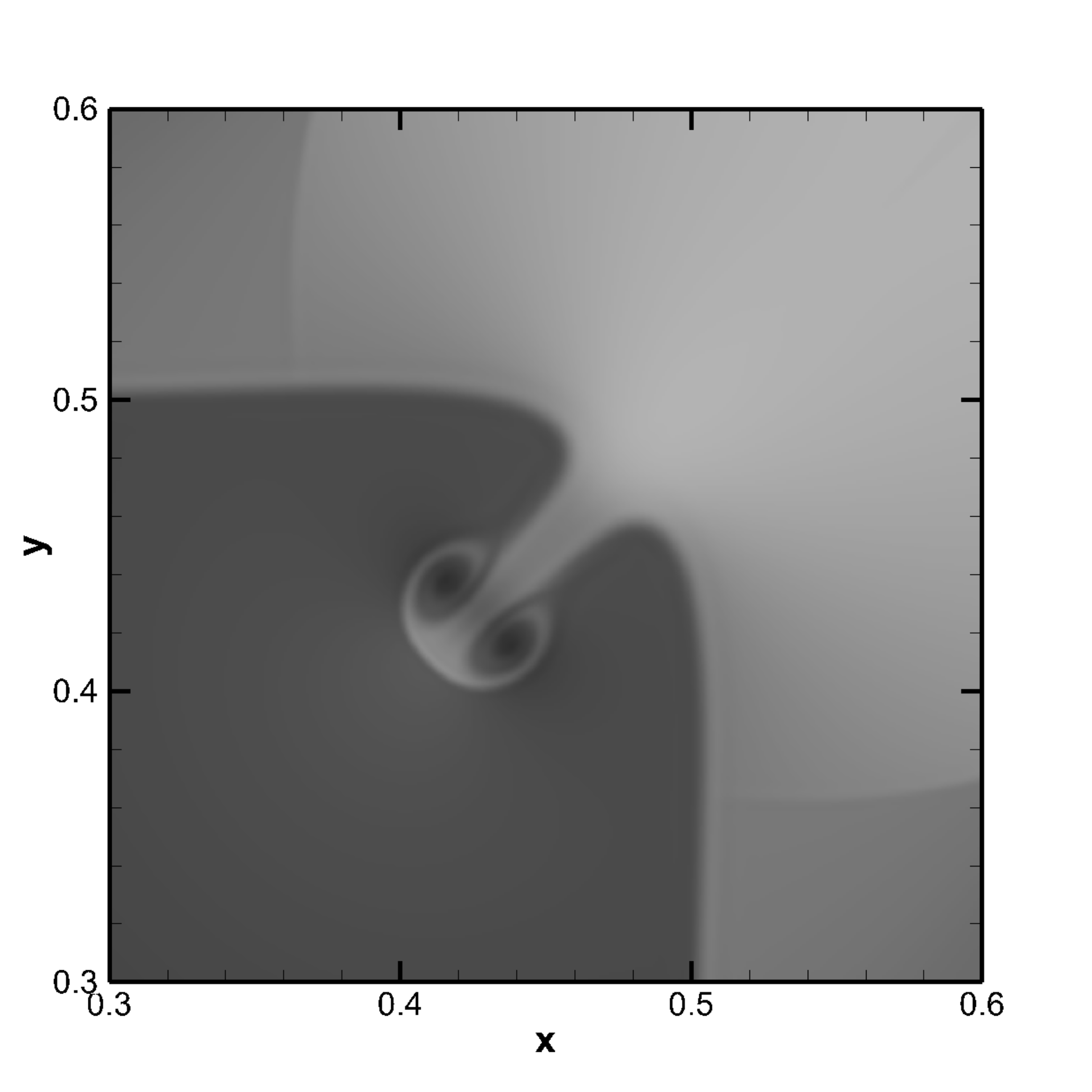}}
  \subfigure[Present]{
  \includegraphics[width=0.45\textwidth]{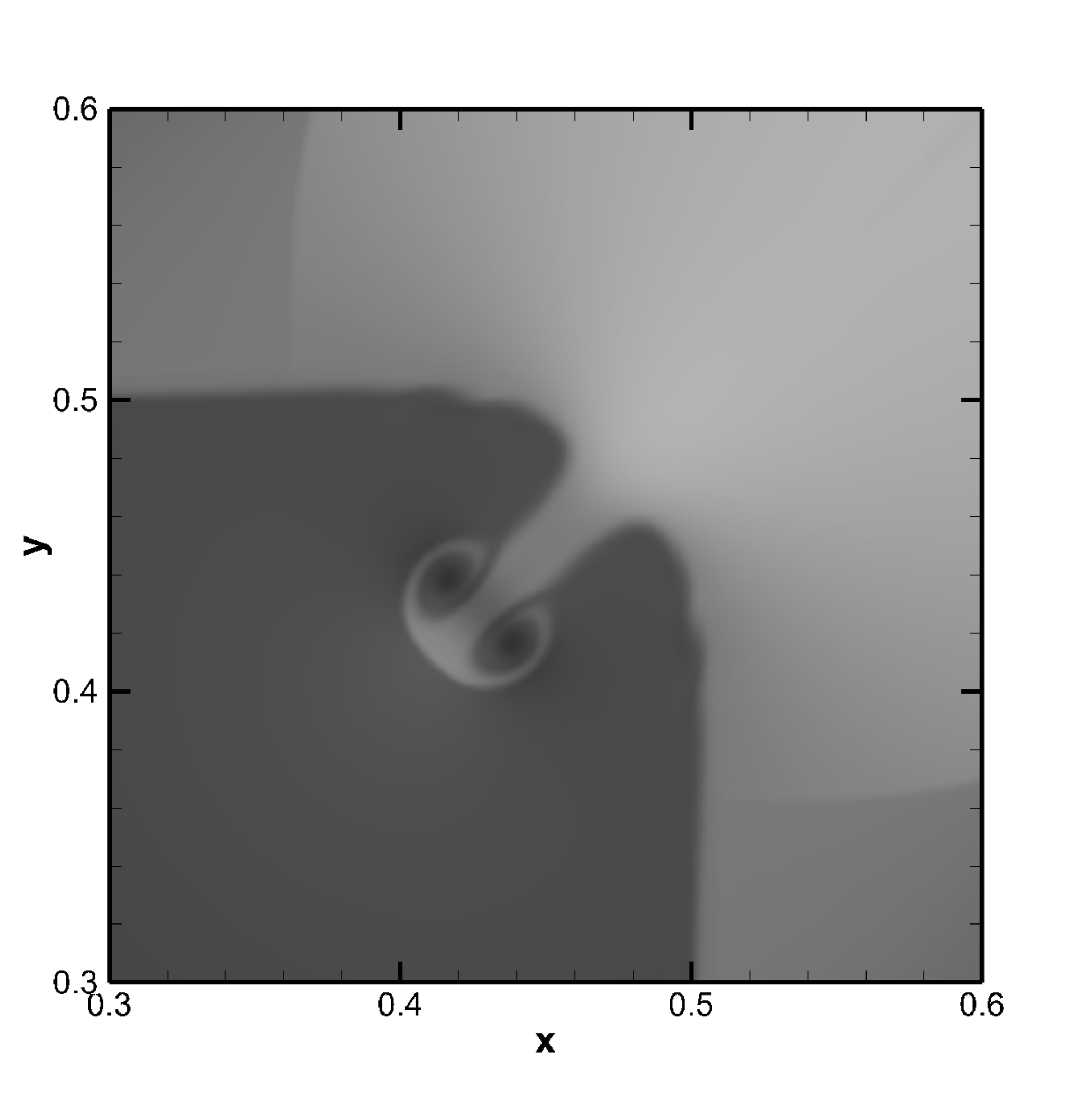}}
  \caption{Density contours for 2D Riemann problem with initial conditions Eq.\eqref{eq:4.13} on a grid of $1024 \times 1024$}\label{fig:9}
\end{center}
\end{figure}

\begin{figure}[H]
  \begin{center}
  \subfigure[WENO-Z]{
  \includegraphics[width=0.45\textwidth]{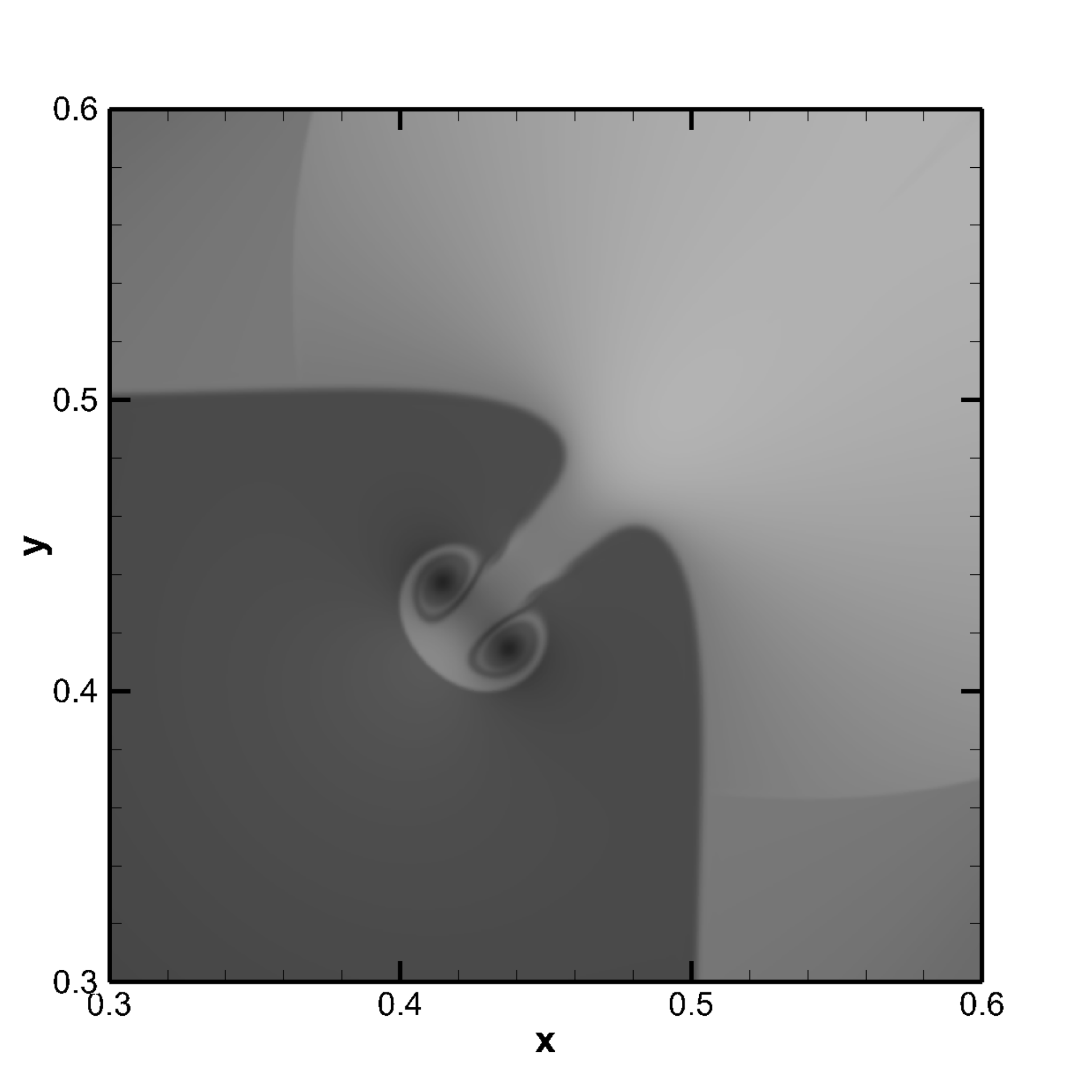}}
  \subfigure[TENO5]{
  \includegraphics[width=0.45\textwidth]{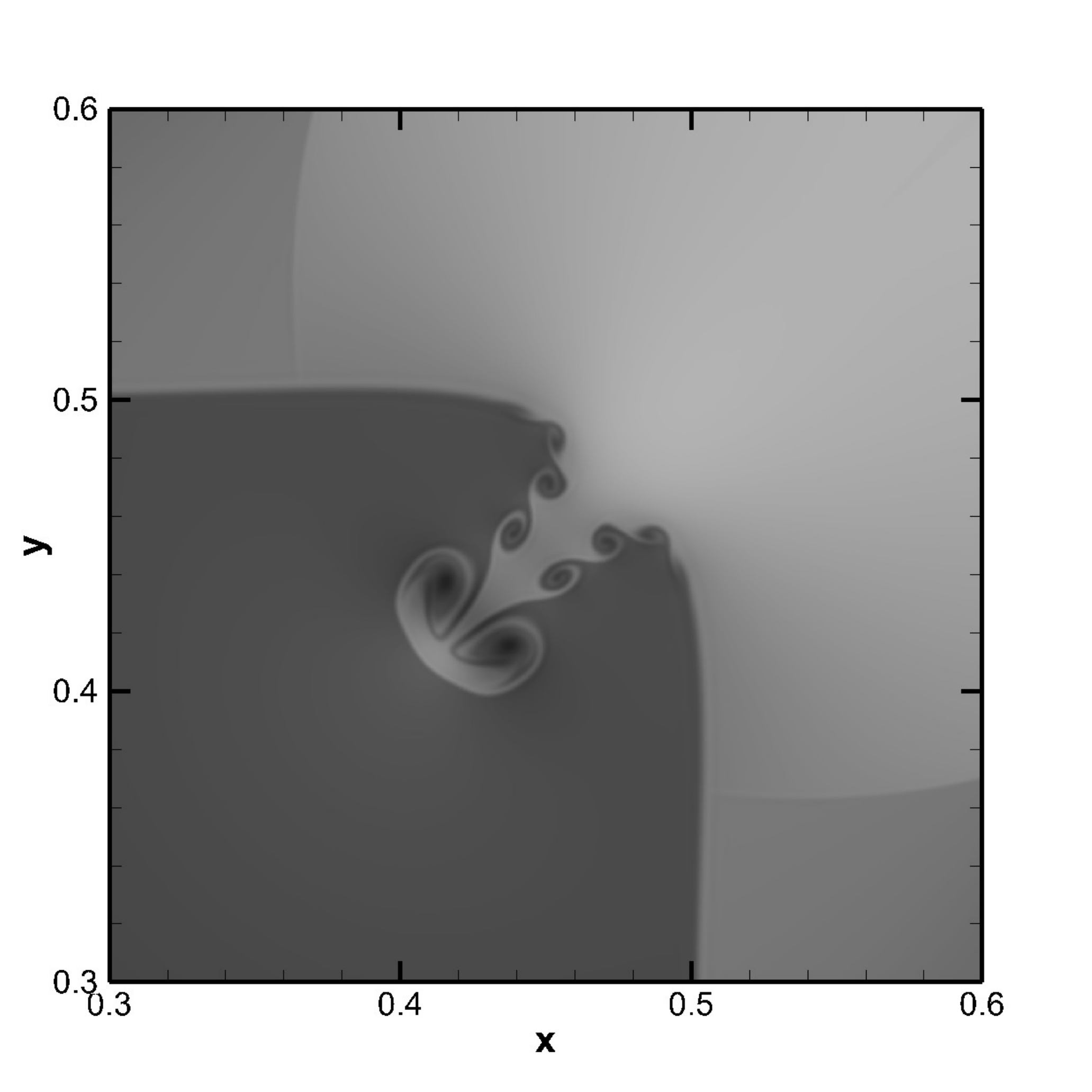}}
  \subfigure[TENO5-A]{
  \includegraphics[width=0.45\textwidth]{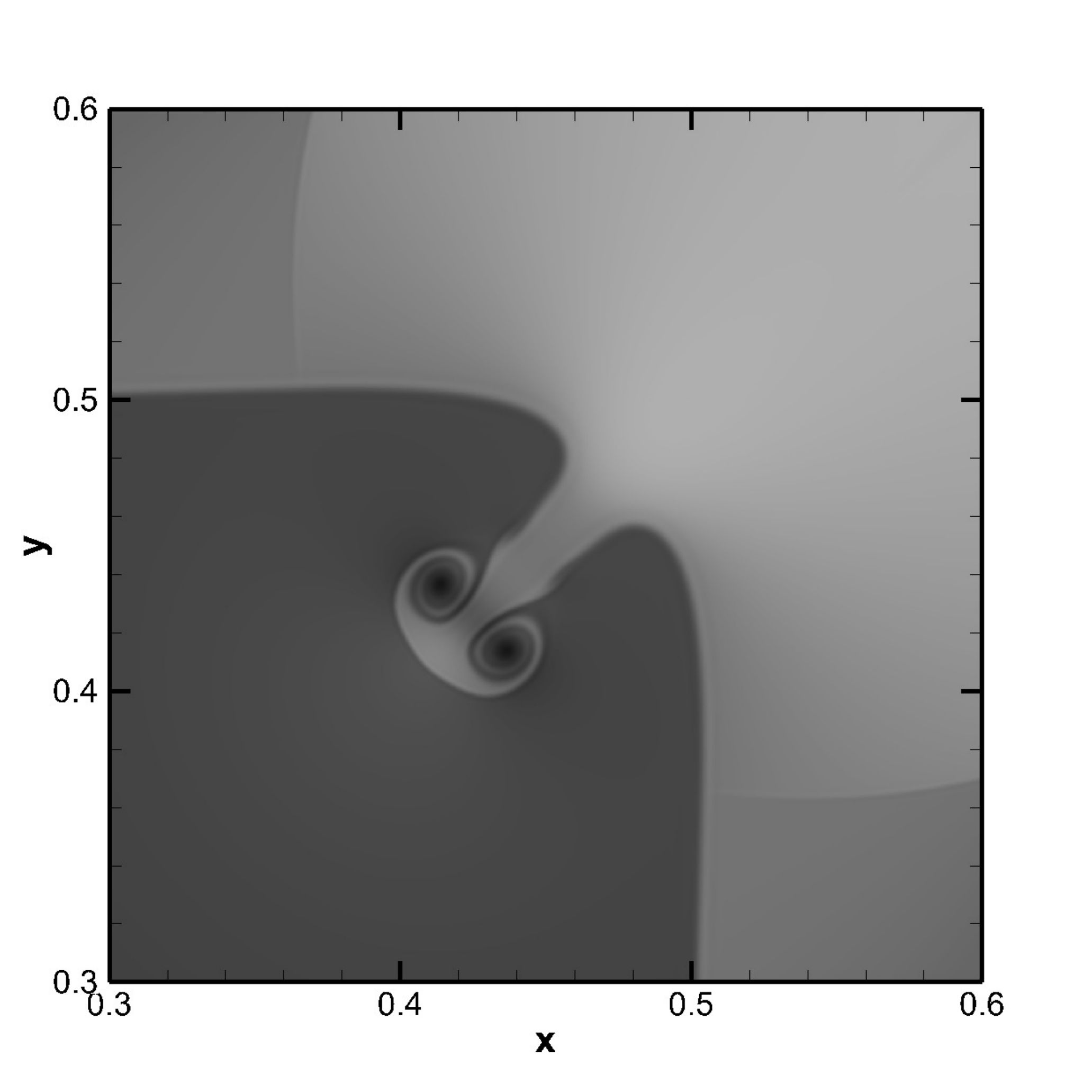}}
  \subfigure[Present]{
  \includegraphics[width=0.45\textwidth]{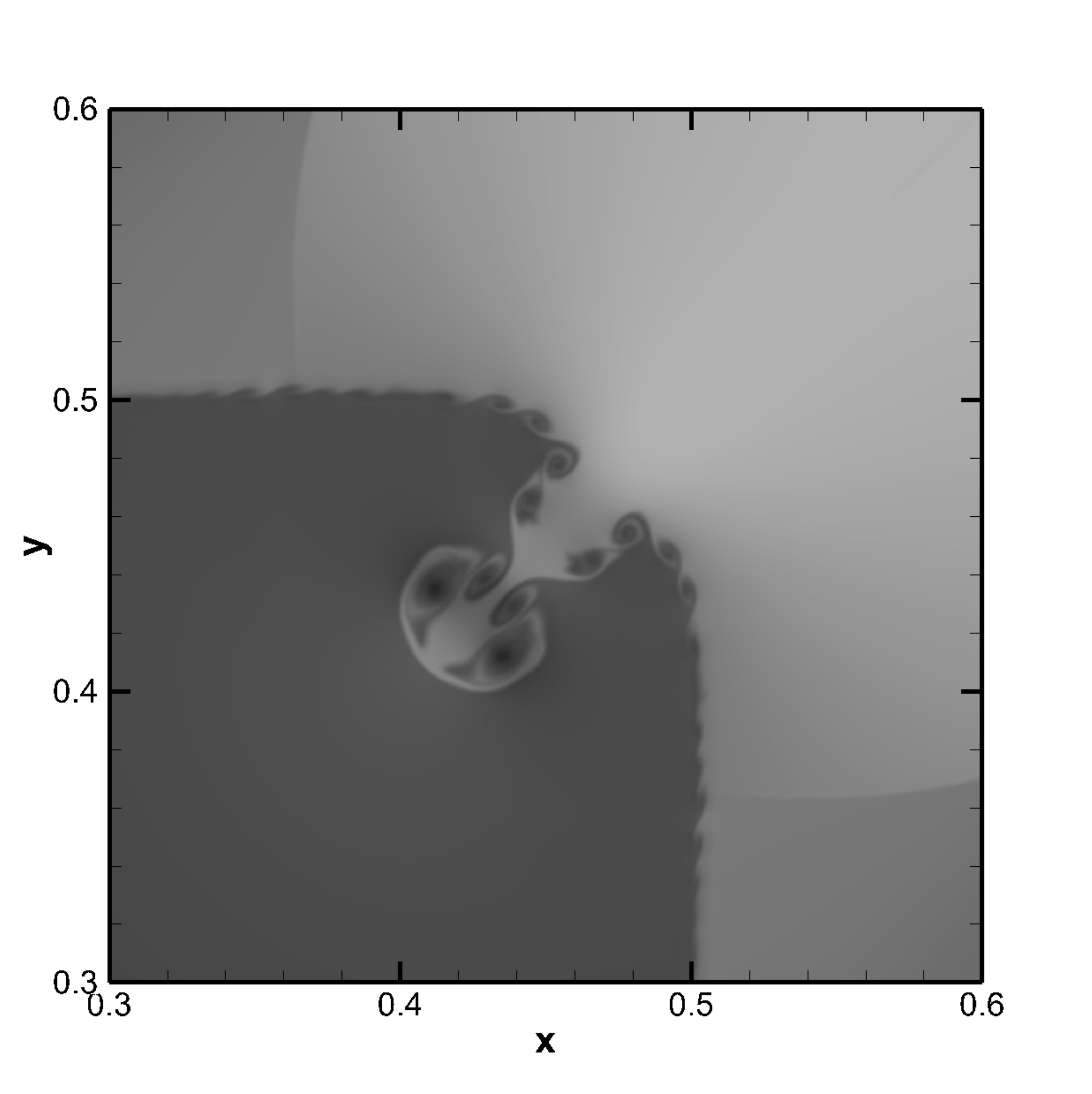}}
  \caption{Density contours for 2D Riemann problem with initial conditions Eq.\eqref{eq:4.13} on a grid of $2048 \times 2048$}\label{fig:11}
\end{center}
\end{figure}

Density contours illustrated in Fig.\ref{fig:9} reveal that although none of the tested schemes reproduce the small structures generated by the KH instability, the proposed scheme is the least dissipative method as some of the instability structures are resolved along the contact line. Density contours on the finer mesh are shown in Fig.\ref{fig:11}. The richness of the small structures resolved by the presented method implies its lower dissipation over the other schemes.

\subsubsection{Double Mach reflection}\label{sec4.3.3}
The double mach reflection test is a mimic of the planar shock reflection in the air from wedges. It is a widely used benchmark to test the ability of shock capturing as well as the small scale structure resolution of a certain scheme. In the present simulation, the computation domain is taken as $[0,4]\times[0,1]$. The lower boundary is set to be a reflecting wall starting from $x=\frac{1}{6}$. At $t=0$, a right-moving $60^\circ$ inclined Mach 10 shock is positioned at $(\frac{1}{6} ,0)$. The upper boundary is set to describe the exact motion of the Mach $10$ shock. The left boundary at $x=0$ is assigned with post-shock values. Zero gradient outflow condition is set at $x=4$. Readers may refer to \cite{woodward1984numerical,kemm2016proper} for detailed descriptions of the double Mach reflection problem. A uniform grid is used with $\Delta x=\Delta y=\frac{1}{256}$. 
\begin{figure}[H]
  \begin{center}
  \subfigure[WENO-Z]{
  \includegraphics[width=1.0\textwidth]{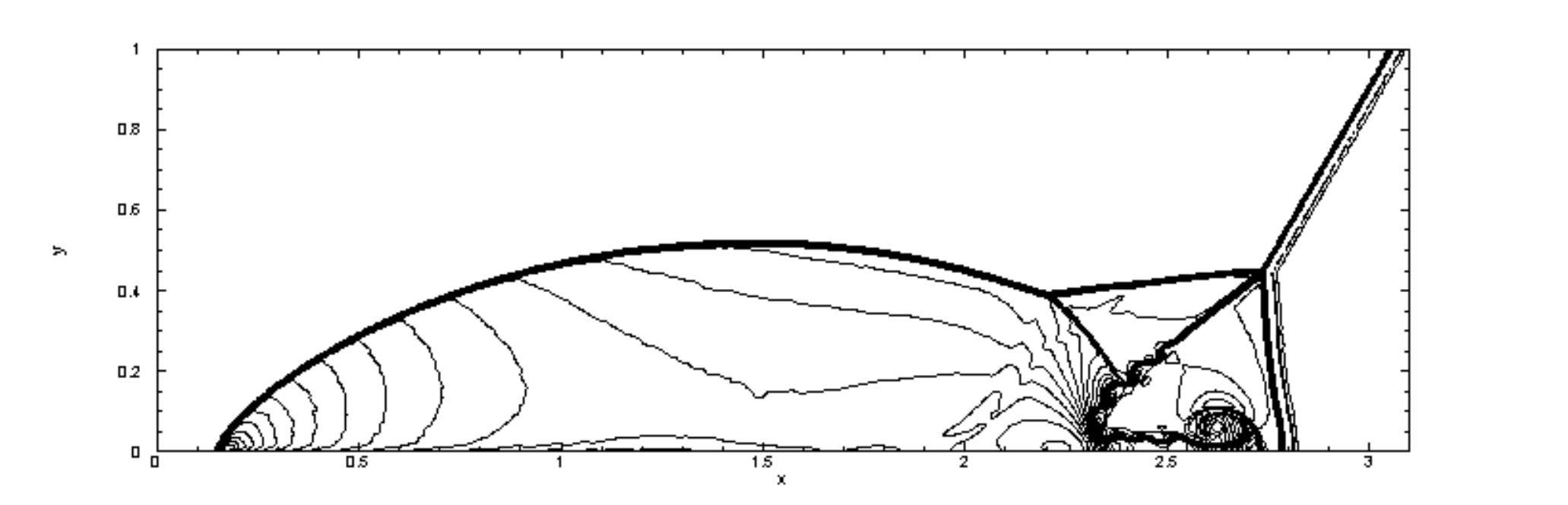}}
  \subfigure[TENO5]{
  \includegraphics[width=1.0\textwidth]{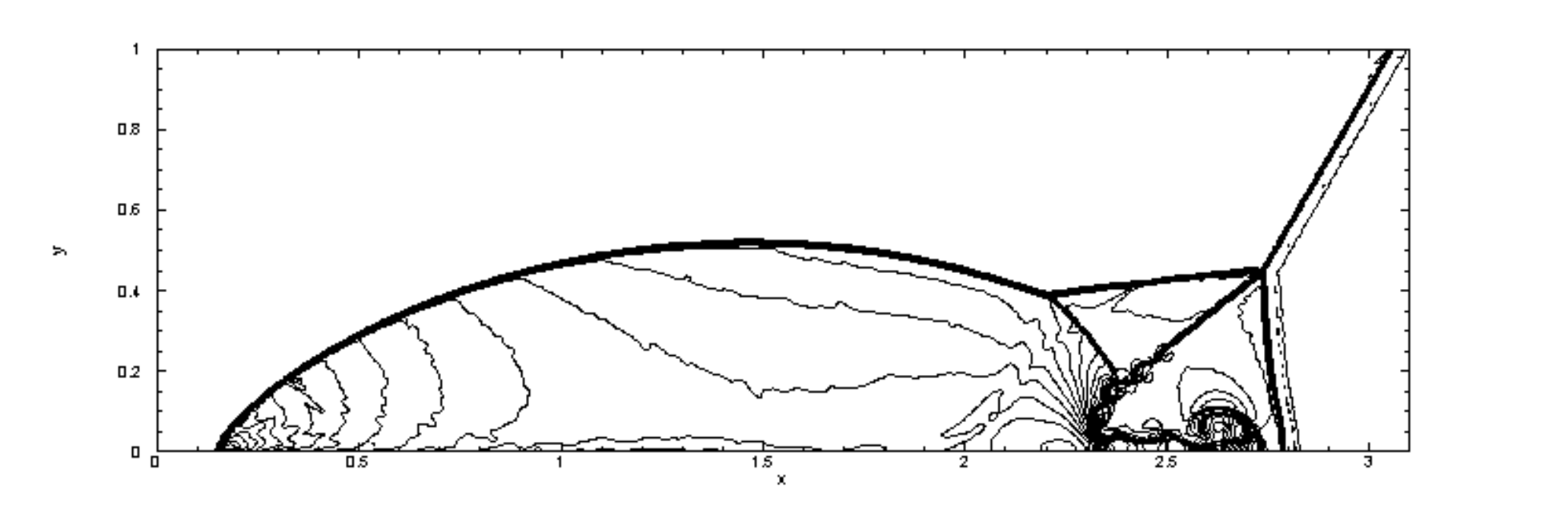}}
\end{center}
\end{figure}

\begin{figure}[H]
  \begin{center}
  \subfigure[TENO5-A]{
  \includegraphics[width=1.0\textwidth]{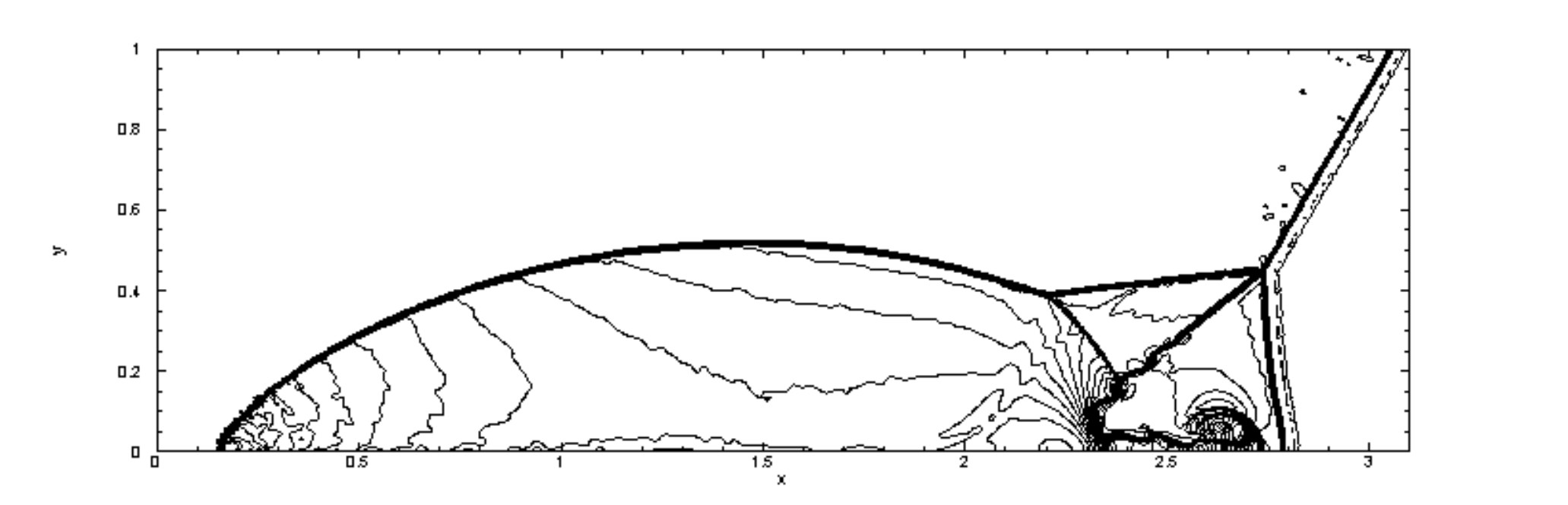}}
  \subfigure[Present]{
  \includegraphics[width=1.0\textwidth]{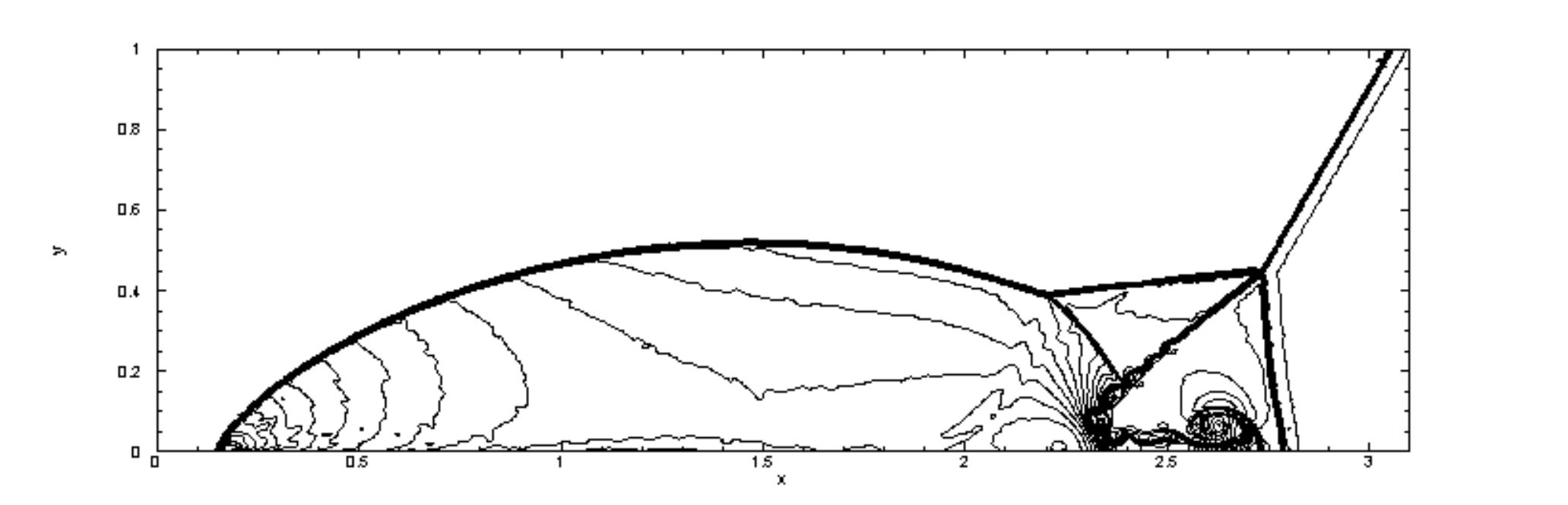}}
  \caption{Density contours of the Double Mach Reflection problem, ranging from $\rho=1.4$ to 21 with 45 equally separated levels.}
\label{fig:12}
\end{center}
\end{figure}

\begin{figure}[H]
  \begin{center}
  \subfigure[WENO-Z]{
  \includegraphics[width=0.45\textwidth]{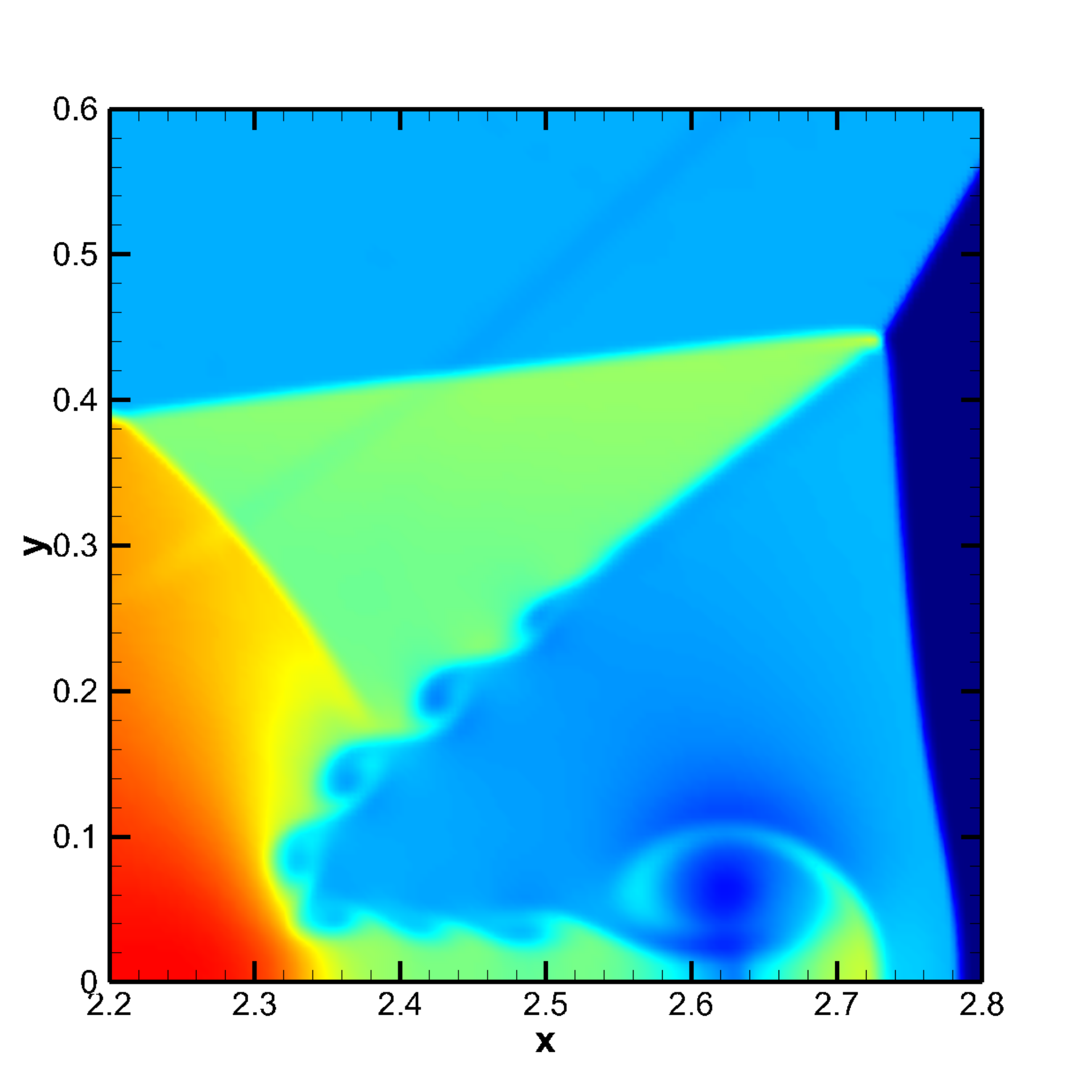}}
  \subfigure[TENO5]{
  \includegraphics[width=0.45\textwidth]{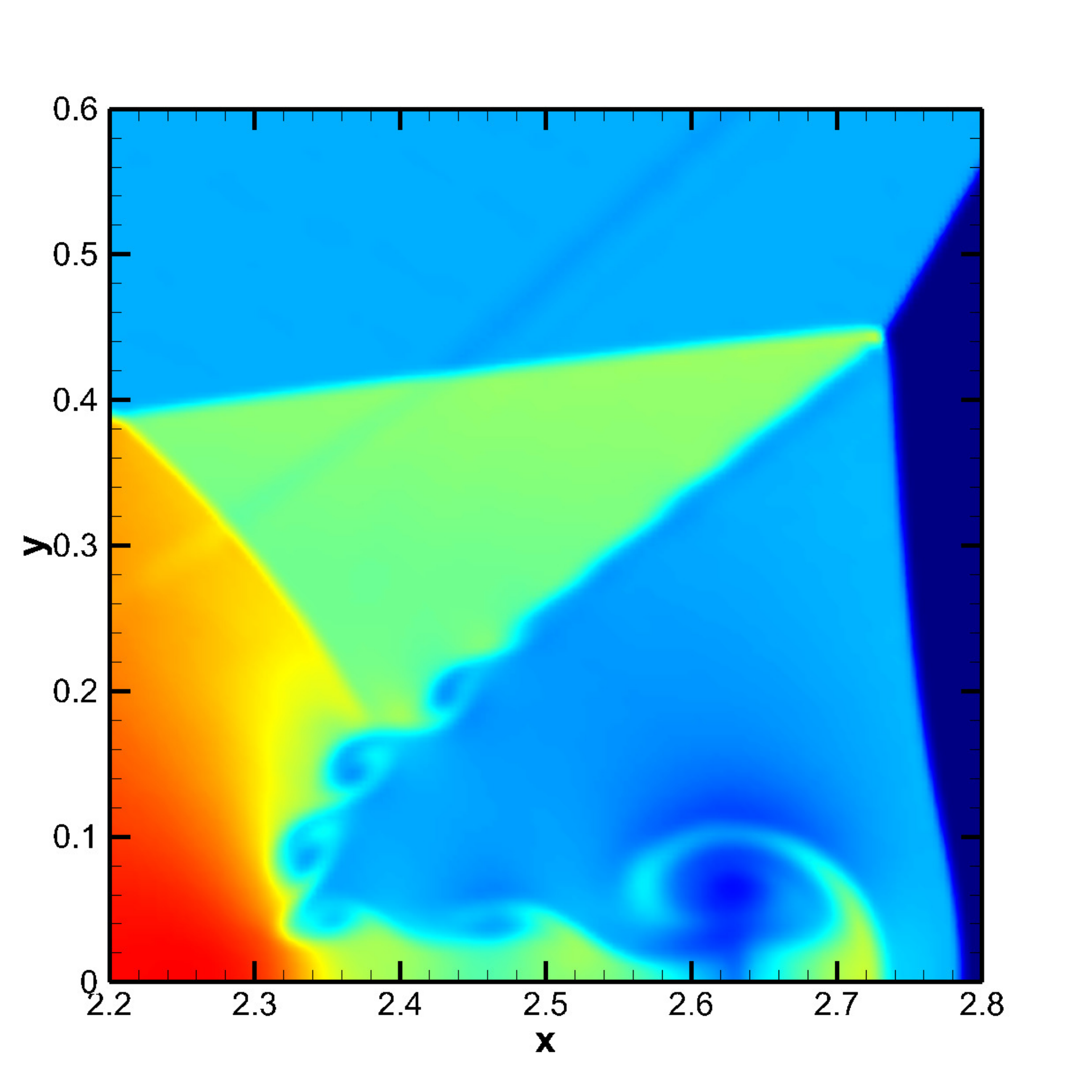}}
  \subfigure[TENO5-A]{
  \includegraphics[width=0.45\textwidth]{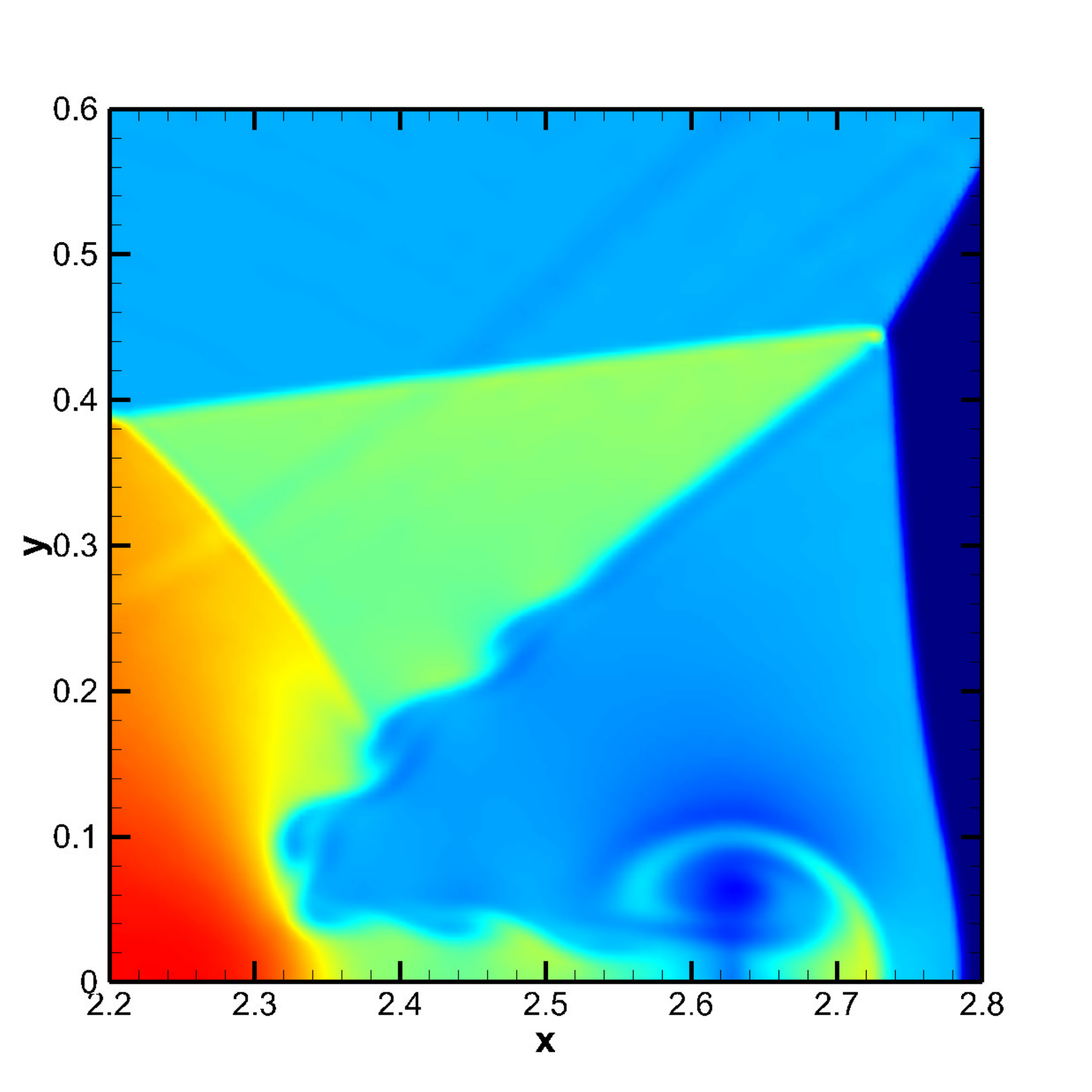}}
  \subfigure[Present]{
  \includegraphics[width=0.45\textwidth]{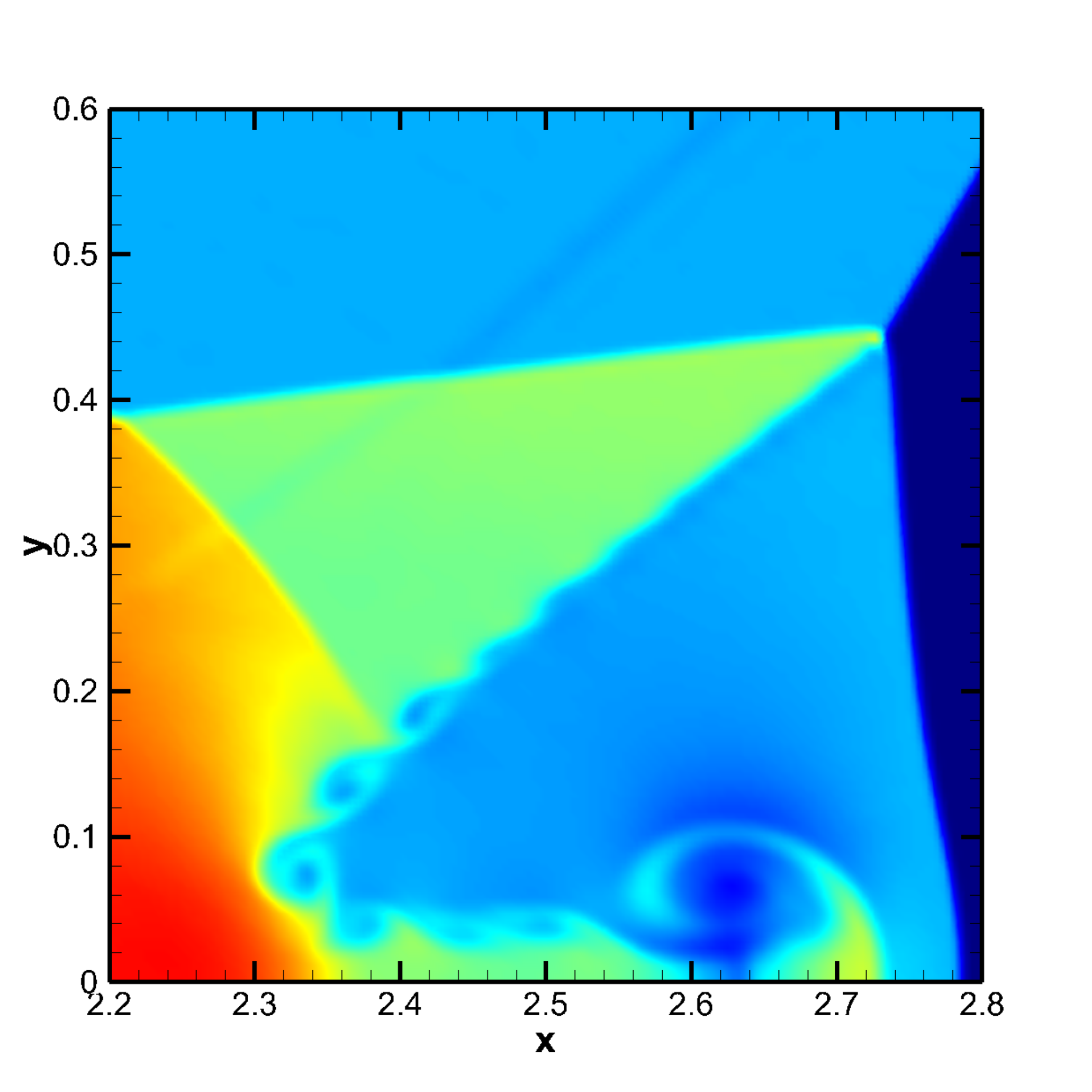}}
  \caption{Zoom-in view of the roll-up region.}
\label{fig:13}
\end{center}
\end{figure}

For this case, the parameters of both of TENO5 and TENO5-A are adjusted to avoid blow-up. The parameter of TENO5 is set to be:
$$
C_T=10^{-5}
$$
The parameters of TENO5-A are set as:
$$
a_1 = 10.5, \quad a_2 = 5.5, \quad C_r=0.3, \quad \xi = 10^{-3}.
$$
Fig.\ref{fig:12} shows the density contours of different methods at $t=0.2$. All methods capture discontinuities. Density contours of the roll-up region of different methods are shown in Fig.\ref{fig:13}. Compared to the other scheme, the presented method resolves the KH instability structures with lower dissipation.

\subsubsection{computational efficiency}\label{sec4.3.4}
The computational time of each scheme for different 2D cases are given in Tab.\ref{tab:2}. All tests are computed on the same desktop workstation. TENO5-LAD needs about $5\%$ more time than TENO5 while TENO5-A needs about $30\%$ more computational.

\begin{table}[H]
  \caption{Averaged computational time (in second) of a single time step of each scheme for different 2D cases, normalized values with respect to the computational time of TENO5 are given in the brackets.}\label{tab:2}
  \begin{center}\footnotesize
  \begin{tabular}{cccccc}
  \toprule
  \multicolumn{1}{c}{Case}    &
  \multicolumn{1}{c}{Grid number}  &
  \multicolumn{1}{c}{WENO-Z}  &
  \multicolumn{1}{c}{TENO5}   &
  \multicolumn{1}{c}{TENO5-A} &
  \multicolumn{1}{c}{Present} \\
  \midrule
  Rayleigh–Taylor instability  & 128$\times$512  & 0.14(0.93) & 0.15(1.0) & 0.2(1.33) & 0.16(1.07) \\
  \quad & 256$\times$1024 & 0.59(0.88) & 0.67(1.0) & 0.88(1.31) & 0.70(1.04) \\
  \quad & \quad & \quad & \quad & \quad & \quad   \\
  2D Riemann Problem Case 1 & 1024$\times$1024    & 2.57(0.88) & 2.89(1.0) & 3.71(1.28) & 2.99(1.03) \\
  \quad & \quad & \quad & \quad & \quad & \quad   \\
  2D Riemann Problem Case 2 & 1024$\times$1024    & 2.56(0.90) & 2.85(1.0) & 3.68(1.29) & 2.99(1.05) \\
  \quad & 2048$\times$2048 & 11.9(0.91) & 13.1(1.0) & 17.8(1.36) & 13.7(1.04)   \\
  \quad & \quad & \quad & \quad & \quad & \quad   \\
  Double Mach Reflection & 1024$\times$256    & 0.59(0.89) & 0.66(1.0) & 0.87(1.31) & 0.69(1.05)   \\
  \bottomrule
  \end{tabular}
  \end{center}
\end{table}

\section{Conclusion}\label{sec5}
In this paper, an efficient target ENO scheme, the TENO5-LAD scheme, is proposed for compressible flow simulation. By utilizing a novel adaptive method, the cutoff parameter $C_T$ is dynamically adjusted according to the smoothness of the reconstruction stencil. Numerical results show that the presented method maintains both of the ENO property and the low dissipation property of the TENO scheme at lower extra computational cost. Due to the novel adaptive method, the presented method is more robust than the TENO5 scheme and the TENO5-A scheme. As fewer tunable parameters are introduced, the new scheme requires lees of the time consuming tuning process and the pre-chosen value of the parameters can be applied to a wide range of cases. Furthermore, as only the most essential ingredients of TENO are required for the adaptive process, the presented method can be directly extended to higher order TENO schemes. 

\section*{Acknowledgement}
The first author is partially supported by the Youth Program of National Natural Science Foundation of China under Grant No.11902326 and the LHD Youth Innovation Fund under Grant No.LHD2019CX05. The second author is partially supported by the National Natural Science Foundation of China under Grants Nos.11872067 and 91852203, NKRDPC No.2016YFA0401200, and SCP No.TZ2016002.
\bibliography{mybibfile}
\end{document}